\documentclass[12pt]{article}
\pdfoutput=1

\usepackage{verbatim}
\usepackage{jheppub}
\usepackage{hyperref}
\usepackage{natbib}
\usepackage{url}

\usepackage{color}
\let\ifpdf\relax
\let\normalcolor\relax  

\def\be{\begin{equation}}
\def\ee{\end{equation}}    
\def\ba{\begin{eqnarray}}
\def\ea{\end{eqnarray}}

\def\lsim{\mbox{\raisebox{-.6ex}{~$\stackrel{<}{\sim}$~}}}

\def\nn{\nonumber \\}

\def\veck{\vec{k}}

\newcommand{\sarah}[1]{\textcolor{red}{[{\bf Sarah}: #1]}}

\title{Large non-Gaussian Halo Bias from Single Field Inflation}
\author{Ivan Agullo,}
\author{Sarah Shandera}
\affiliation{ Institute for Gravitation and the Cosmos, The Pennsylvania State University, University Park, PA 16802, USA}
\emailAdd{agullo@gravity.psu.edu}
\emailAdd{shandera@gravity.psu.edu}

\abstract{We calculate Large Scale Structure observables for non-Gaussianity arising from non-Bunch-Davies initial states in single field inflation. These scenarios can have substantial primordial non-Gaussianity from squeezed (but observable) momentum configurations. They generate a term in the halo bias that may be more strongly scale-dependent than the contribution from the local ansatz. We also discuss theoretical considerations required to generate an observable signature.}

\begin{document}
\maketitle

\section{Introduction}

The study of the primordial cosmological perturbations has recently shifted focus toward a careful analysis of statistics of the fluctuations beyond the power spectrum. The higher order statistics, collectively called non-Gaussianity, are an extremely rich source of new information about the origin of the cosmic inhomogeneities. To make the most of this information we must both measure the effects of any primordial non-Gaussianity and understand the implications of the result for theories of the very early universe. 

The goal of this paper is to work out new observational consequences of a conceptually important aspect of {\it any} inflationary scenario: the quantum initial state. The choice of initial state, and how observationally relevant it is, has long been a source of debate among inflationary theorists \cite{Mottola:1984ar,Martin:2000xs,Kaloper:2002uj,Schalm:2004xg}. Several years ago, interest in the idea was driven by the possibility of seeing evidence of some high energy scale $M$ in corrections to the power spectrum \cite{Easther:2001fi,Easther:2001fz,Easther:2002xe,Shiu:2002kg,Danielsson:2002kx,Kaloper:2002cs,Martin:2003kp,Nitti:2005ym,Greene:2005aj}. Generically $M>H$, where $H$ is the Hubble rate during inflation and $M$ could be the string scale, for example. Any corrections to the power spectrum depend on powers of the small ratio $H/M$. For a general initial state, scenarios that are not already observationally ruled out can at most add a high frequency, small amplitude oscillation on top of the nearly scale-invariant, monotonic power spectrum. In addition, an observable signal seems to require a fine-tuning in the time that one specifies the initial state compared to the time when modes observable in the Cosmic Microwave Background (CMB) exit the horizon. So, although intriguing, the possibility of observing generic initial states was not widely considered likely to be observationally interesting. 

However, the study of non-Gaussianity has shown that higher order correlation functions are more sensitive to physics at scales $M>H$ than the power spectrum is. The amplitude of non-Gaussianity generically increases the closer to $H$ the scale $M$ is; conversely, for fixed $H$, higher order correlations become unobserveably small as $M$ approaches the Planck scale $M_P$. Non-Gaussianity then offer a more powerful tool to study the physical consequences of the initial conditions for inflation. 

The motivation for considering a generic initial state is the same as the motivation for considering any other kind of non-Gaussianity: we do not know the particle physics of inflation, or how long inflation lasted or what came before. At best we might parametrize our ignorance in terms of an effective description at scale $M$. In that case, it is natural to expect modifications to the initial quantum state {\it together with} new interaction terms in the effective Lagrangian of the inflaton. Some scenarios may appear more fine-tuned than others, but fortunately we have observations to act as a counterpoint to the theoretical prejudices of the moment. 

Generalizations of the initial state, through gravitational interactions alone, produce a primordial three point correlation function, or bispectrum, with a large amplitude in squeezed triangle configurations where one of the momenta is much smaller than the others ($k_3\ll k_2,k_1$) \cite{Agullo:2010ws,Agullo:2011xv}. This type of bispectrum has a very significant effect in the power spectrum of gravitationally bound objects like galaxies and galaxy clusters. In anticipation of further improved constraints and a generalized analysis of data from future surveys \cite{SDSS, BOSS, HETDEX, LSST, DES}, we work out the observational consequences for Large Scale Structure from primordial non-Gaussianity arising from a generalized initial state. Existing data has not yet been analyzed with the bispectrum we study here in mind, but we will show that in principle a generalized initial state could already be well constrained with the existing measurements of the halo bias \cite{Dalal:2007cu,Slosar:2008hx}. Previous work on the bispectrum, largely focused on the CMB, can be found in \cite{Chen:2006nt, Holman:2007na,Meerburg:2009ys,Meerburg:2009fi,Meerburg:2010rp,Agullo:2010ws,Ganc:2011dy,Chialva:2011hc,Ashoorioon:2010xg}. A complementary analysis to ours of the effects of a generalized initial state on the CMB and large scale structure has been simultaneously completed by J. Ganc and E. Komatsu \cite{GancKomatsu}, and some of their early results were previously presented by E. Komatsu at the `Pre-Planckian Inflation' conference \cite{KomatsuTalk}.

The main points we will stress in this paper are:
\begin{itemize}
\item The bispectrum arising from a generic initial state is an example of non-Gaussianity that is large in squeezed momentum configurations ($k_3\ll k_1,k_2$, all scales observable today), and is {\it  single field} in origin. This is unexpected from the point of view of the consistency relation \cite{Maldacena:2002vr, Creminelli:2004yq,Creminelli:2011rh}, but not necessarily inconsistent.

\item This bispectrum can lead to a term in the halo bias that is more strongly scale dependent than the analogous contribution from local ansatz \cite{Salopek:1990jq, Verde:1999ij,Komatsu:2001rj} non-Gaussianity.

\item The amplitude of the non-Gaussian bias receives the most significant contributions from squeezed, nearly collinear momentum configurations (sometimes called `elongated' in the literature) as well as subdominant contributions from  squeezed-isosceles configurations. 
\end{itemize}

These points are explained in detail in the rest of the paper. In Section \ref{sec:GIS} we introduce a phenomenological form of the bispectrum and study the main characteristics of its shape. We review the theoretical motivation for the bispectrum arising from non-vacuum initial states, and consider some illustrative examples. However, the theory discussion is self-contained and can be skipped by readers interested in just the phenomenological consequences for Large Scale Structure (LSS).  In Section \ref{sec:LSS} we compute some observational signatures in Large Scale Structure arising from this bispectrum. We conclude with the implications for parameterizing future LSS constraints on primordial non-Gaussianity.

\section{The Generalized Initial State (GIS) Bispectrum\label{GIS}}
\label{sec:GIS}
The bispectrum, $B_{\zeta}$, for primordial curvature perturbations $\zeta$ is defined in terms of the three point correlation function in momentum space
\ba \label{threepoint}
\langle \zeta_{\vec{k}_1}\zeta_{\vec{k}_2}\zeta_{\vec{k}_3}\rangle &=&(2\pi)^3\delta^3_D(\vec{k}_1+\vec{k}_2+\vec{k}_3)\;B_{\zeta}(\veck_1,\veck_2,\veck_3) \, .
\ea
In a similar way, the power spectrum is defined in terms of the two point function by
\ba
\langle \zeta_{\vec{k}_1}\zeta_{\vec{k}_2}\rangle &=&(2\pi)^3\delta^3_D(\vec{k}_1+\vec{k}_2)\;P_{\zeta}(k_1) \, .
\ea
It is convenient to define the dimensionless power spectrum $\mathcal{P}_{\zeta}(k)\equiv\frac{k^3}{2\pi^2} P_{\zeta}(k)$. 
The bispectrum we analyze in this paper arises in models of inflation in which the quantum state of comoving curvature perturbations $\zeta$ is an excited state compared to the Bunch-Davies vacuum. This bispectrum, which we label GIS after its origin in a Generalized Initial State for inflation, can be written as
\ba \label{BGIS} 
&& \hspace{-1cm} B_{\rm GIS}(k_1,k_2,k_3)= \mathcal{B}_{{\rm GIS}} \ P_{\zeta}(k_1)P_{\zeta}(k_2) \ \frac{k_1^2k_2^2}{k_3^3}  \ \times \\\nonumber 
&\times&\,  \, \textrm{Re}\left[f_t\frac{1-e^{ik_t/k_*}}{k_t}+f_{1}\frac{1-e^{i\tilde{k}_1/k_*}}{\tilde{k}_1}+f_{2}\frac{1-e^{i\tilde{k}_2/k_*}}{\tilde{k}_2}+f_{3}\frac{1-e^{i\tilde{k}_3/k_*}}{\tilde{k}_3} \right]   +\textrm{2 perm.}\, ,\ea
where $k_t=k_1+k_2+k_3$, $\tilde{k}_i=k_t-2 k_i$, and $\mathcal{B}_{{\rm GIS}}$ is a coefficient parameterizing its amplitude. The functions $f_{i}$ may have dependence on the momenta, and generically have non-vanishing real and imaginary parts. For comparison, the well studied local ansatz \cite{Salopek:1990jq, Verde:1999ij,Komatsu:2001rj} is
\ba \label{local}
B_{\rm local}(k_1,k_2,k_3) = \mathcal{B}_{{\rm local}} \ [P_{\zeta}(k_1)P_{\zeta}(k_2)  \, + \, \textrm{2 perm.}]\, ,\ea
with $\mathcal{B}_{local}\equiv\frac{6}{5}f_{NL}$. The next subsections discuss important features of the GIS bispectrum, including theoretical characteristics and observational constraints on the parameters in $B_{{\rm GIS}}$. Subsection \ref{sec:TheoryLight} contains a brief list of the most important points connecting the phenomenological ansatz above to scenarios with a modified initial state. Subsection \ref{sec:ShapeAnalyze} analyzes the shape $B_{{\rm GIS}}$. Finally, Subsection \ref{sec:examples} discusses some additional details of physically and observationally reasonable initial states, but can be skipped by an observationally minded reader.

\subsection{Lightening theory review}
\label{sec:TheoryLight}
Any complete particle physics model of inflation should specify not only the action for the relevant matter fields and how they couple to gravity, but also the initial conditions both for the classical background spacetime and for the quantum state the fluctuations start in. The quantum initial state is usually taken to be de Sitter invariant vacuum state, the so called Bunch-Davies vacuum \cite{Bunch:1978yq}. This assumption may be too restrictive, because we do not know how long inflation lasted or what expansion history preceded it. Therefore, just as an effective theory should include generic interaction terms in the Lagrangian, it should also allow a generic initial state that is consistent with inflation. 

The bispectrum for a modified initial state was first computed in \cite{Chen:2006nt, Holman:2007na}, but the relevance for the enhancement in the squeezed configuration was first recognized in \cite{Agullo:2010ws} and further analyzed in \cite{Ganc:2011dy}. This was, to our knowledge, the first example of a scenario that is single-field in the usual sense (only one degree of freedom is relevant for the background inflationary expansion and for the power spectrum) but that nonetheless has large non-Gaussianity in the squeezed triangles observable in our universe post-inflation. Small scale features during inflation that generate subhorizon interactions can also lead to a stronger signal in squeezed but observable triangles \cite{Chen:2008wn}. Another mechanism for boosting the squeezed limit was recently found in \cite{Barnaby:2012tk}). The bispectrum has the form given in Eq.(\ref{BGIS}), but when the non-Gaussianity originates from a single field scenario with a Generalized Initial State the following properties hold:

\begin{itemize}

\item {\bf $k_*$ is a long wavelength scale}. 
The scale $k_*$ is related to the value of conformal time $\tau_0$ at which we specify the initial conditions, $|\tau_0|\equiv k_*^{-1}$. The physical condition that the observable modes in our present universe were deeply inside the Hubble radius at time $\tau_0$ translates into
\[\boxed {k_i\gg k_* }\]
 for $i=1,2,3$.
The scale $k_*$ may refer to a genuine transition into inflation or may be the earliest we trust a particular particle description of inflation.

\item {\bf The interactions are gravitational in origin.} The bispectrum above does not depend on the form of the self interactions of the inflaton field (although adding interactions can further enhance the signal). The amplitude of the bispectrum is proportional to the slow-roll parameter $\epsilon=-\frac{\dot{H}}{H^2}$. Including the correct numerical factor, we have 
\[\boxed{\mathcal{B}_{{\rm GIS}}=4\epsilon}\]
\item {\bf The coefficients $f_i$ generically have real and imaginary parts, and are scale dependent.} 
The coefficients $f_i$ encode the information about the initial state (see Section \ref{sec:examples} and the Appendix for explicit expressions and examples). They depend on how the modes $k_1$, $k_2$ and $k_3$ are populated as compared to the Bunch-Davies vacuum, and therefore they are scale dependent. They must decrease for high momentum faster than $(k_*/k)^4$ to ensure acceptable ultra-violet behavior of the initial state. This fall-off may imply small values of all the $f_i$ at observable scales if inflation lasts much longer than the minimum number of  e-folds. However, over a finite range of $k$, the functions $f_i$ may be nearly constant. In fact, the observation of a nearly invariant power spectrum requires the $f_i$, if non-negligible, to be at most weakly scale-dependent (see section \ref{sec:examples} for further details). In addition, for this bispectrum to be observable the $f_i$ should not be dominated by oscillatory terms (which does happen in models with oscillations in the Lagrangian describing the inflationary phase \cite{Flauger:2010ja, Chen:2010bka}). It would be very useful to have a more thorough understanding of both the most generic scale dependence and oscillatory behavior that can appear in the initial state.

\item {\bf The shape contains a piece of the standard slow-roll bispectrum.} Taking the limit in which the initial state approaches the Bunch-Davies vacuum corresponds to considering $f_{1}=f_{2}=f_{3}\to0$ and $f_t\to1$. Then, Eq.(\ref{BGIS})  reduces to
$$ {B(k_1,k_2,k_3)_{GIS\rightarrow BD}=4\epsilon \left[P_{\zeta}(k_1)P_{\zeta}(k_2)\frac{k_1^2k_2^2}{k_3^3k_t}+\textrm{2 perm.}\right]}$$
In the case of Bunch-Davies vacuum there is an additional contribution to the bispectrum (also proportional to slow-roll parameters) that is as important as this one. However, that extra term is not enhanced by changing the initial quantum state, so we neglect it here (see Eq.(\ref{fullbisp}) in the Appendix for the full expression). 
Notice also that in the squeezed limit and the Bunch-Davies vacuum, the term above is $B(k_3\ll k_1\approx k_2)_{GIS\rightarrow BD}\rightarrow2\epsilon P_{\zeta}(k_1)P_{\zeta}(k_3)$ (assuming the power spectrum is nearly scale-invariant). This is part of the familiar single-field slow-roll consistency relation \cite{Maldacena:2002vr}.
\end{itemize}

\subsection{The shape $B_{\rm GIS}(k_1,k_2,k_3)$ and the role of $k_*$ \label{sub:shape}}
\label{sec:ShapeAnalyze}
The GIS bispectrum, $B_{\rm GIS}(k_1,k_2,k_3)$, given in Eq.(\ref{BGIS}), is considerably more complicated than the usual local bispectrum (\ref{local}). Here we study its behavior in terms of momenta and characterize the configurations for which it attains the largest value. Recall that the local ansatz for the bispectrum has a dominant contribution in the squeezed configuration in which $k_3\ll k_2\approx k_1$, given by 
\be
B_{{\rm local}}\rightarrow\frac{12}{5}f_{NL}P(k_1)P(k_3)\;.\ee
We can more clearly illustrate the relative importance of different momenta configurations by factoring out one of the momenta, $k_1$, and studying the behavior of $B_{\rm GIS}$ as a function of the ratios of the other momenta to $k_1$. We define $x_1=1$, $x_2=k_2/k_1$, $x_3=k_3/k_1$, and similarly $x_*=k_*/k_1$ and $\tilde{x}_i=\tilde{k}_i/k_1$. Since the bispectrum is symmetric in the momenta, we can simplify the analysis by restricting to $1\ge x_2\ge x_3$. All other configurations will be related to those by permutations of the momenta. The presence of the Dirac delta in (\ref{threepoint}) forces the three momenta to form a triangle, which translates into the condition $x_3\ge 1- x_2$. We will also require the physically motivated condition $x_i\gg x_*$ (introduced in the previous section). One might impose a similar constraint to consider only momenta observable today, $x_i>x_{\rm min}\gg x_*$.

The novel feature of $B_{\rm GIS}$ is the presence of new contributions proportional to factors of the type $1/\tilde{x}_i$, appearing in the second line of Eq. (\ref{BGIS}). These new factors can be large for some specific momentum configurations, producing a significant enhancement in the bispectrum. For the restricted set of momenta we are considering here, we have the following restrictions

\ba
&&x_t\ \equiv\ \ 1+x_2+x_3 \geq 2\\ \nonumber
&&\tilde{x}_1\, \equiv-1+x_2+x_3\geq0\\ \nonumber
&&\tilde{x}_2\ \ \equiv\ 1-x_2+x_3\geq x_{\rm min}\gg x_*\\ \nonumber
&&\tilde{x}_3\ \ \equiv\ 1+x_2-x_3\geq1
\ea
From this we see that the most important terms are those proportional to $1/{\tilde{x}_1}$ and $1/{\tilde{x}_2}$ in expression (\ref{BGIS}). Let us analyze each of these terms separately.

\begin{figure}[h]
\begin{center}
$\begin{array}{cc}
\includegraphics[width=0.5\textwidth,angle=0]{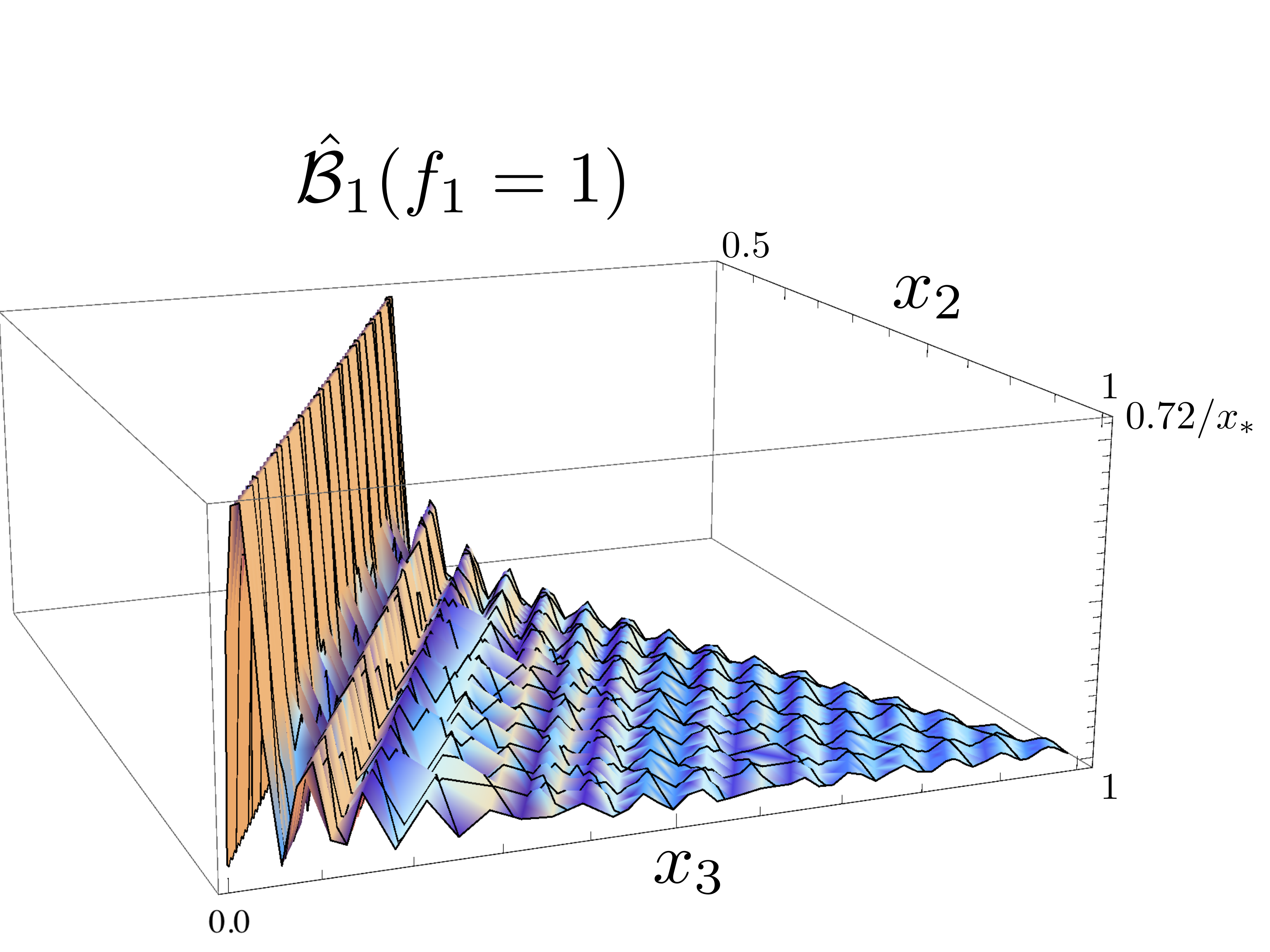} 
&
\includegraphics[width=0.5\textwidth,angle=0]{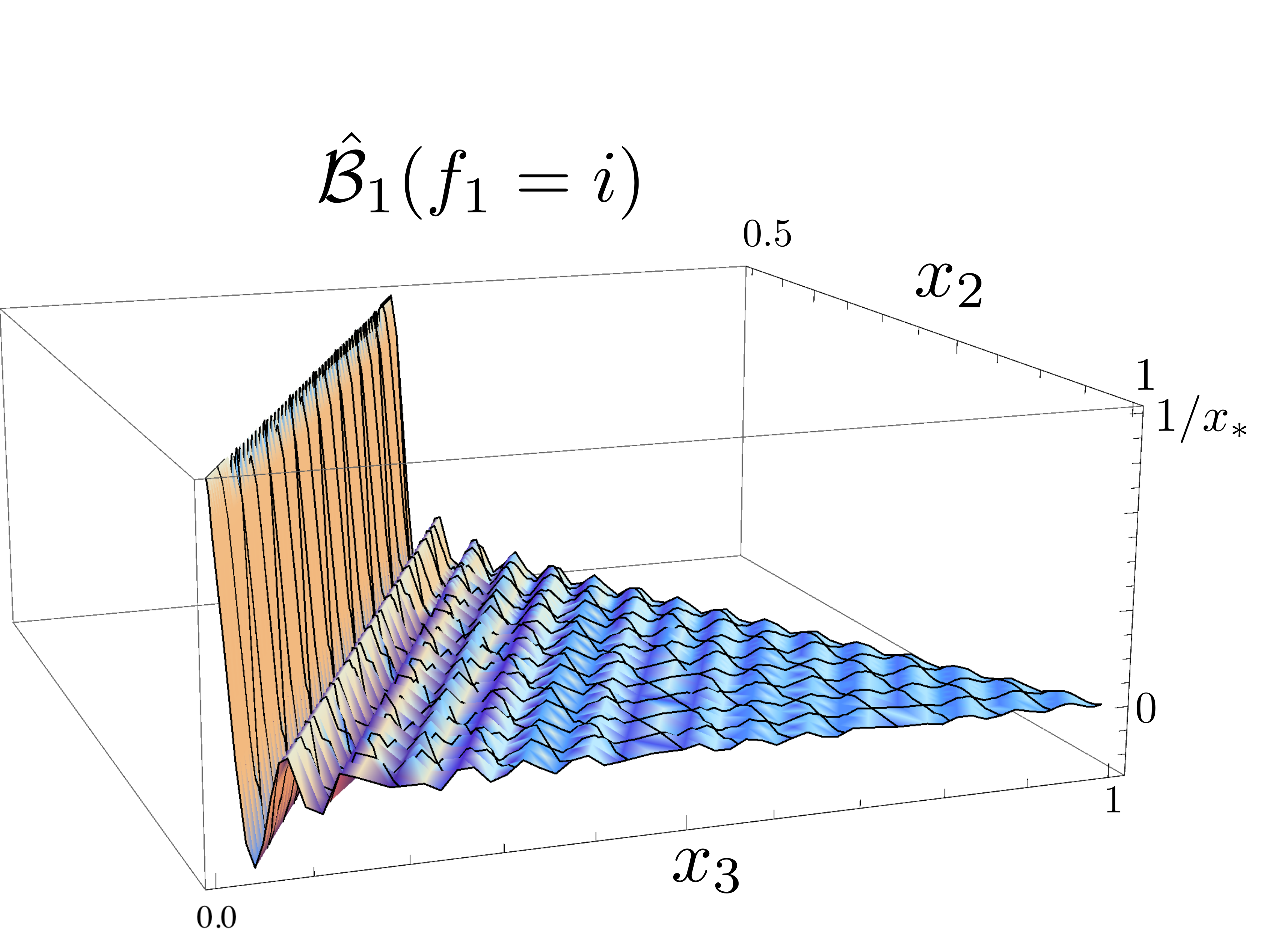} 
\end{array}$
\caption{Contributions from the real (left)  and imaginary (right) part of the $f_1$ to the term shown in (\ref{f1term}). Both are constants in lines $\tilde x_1={\rm constant}$, and attain the maximum value in nearly collinear configurations with $\tilde x_1=2.33 \, \tilde x_*$, and exact collinear configurations $\tilde x_1=0$, respectively.
\label{fig:f1f2}}
\end{center}
\end{figure}

\begin{figure}[h]
\begin{center}
\includegraphics[width=0.7\textwidth,angle=0]{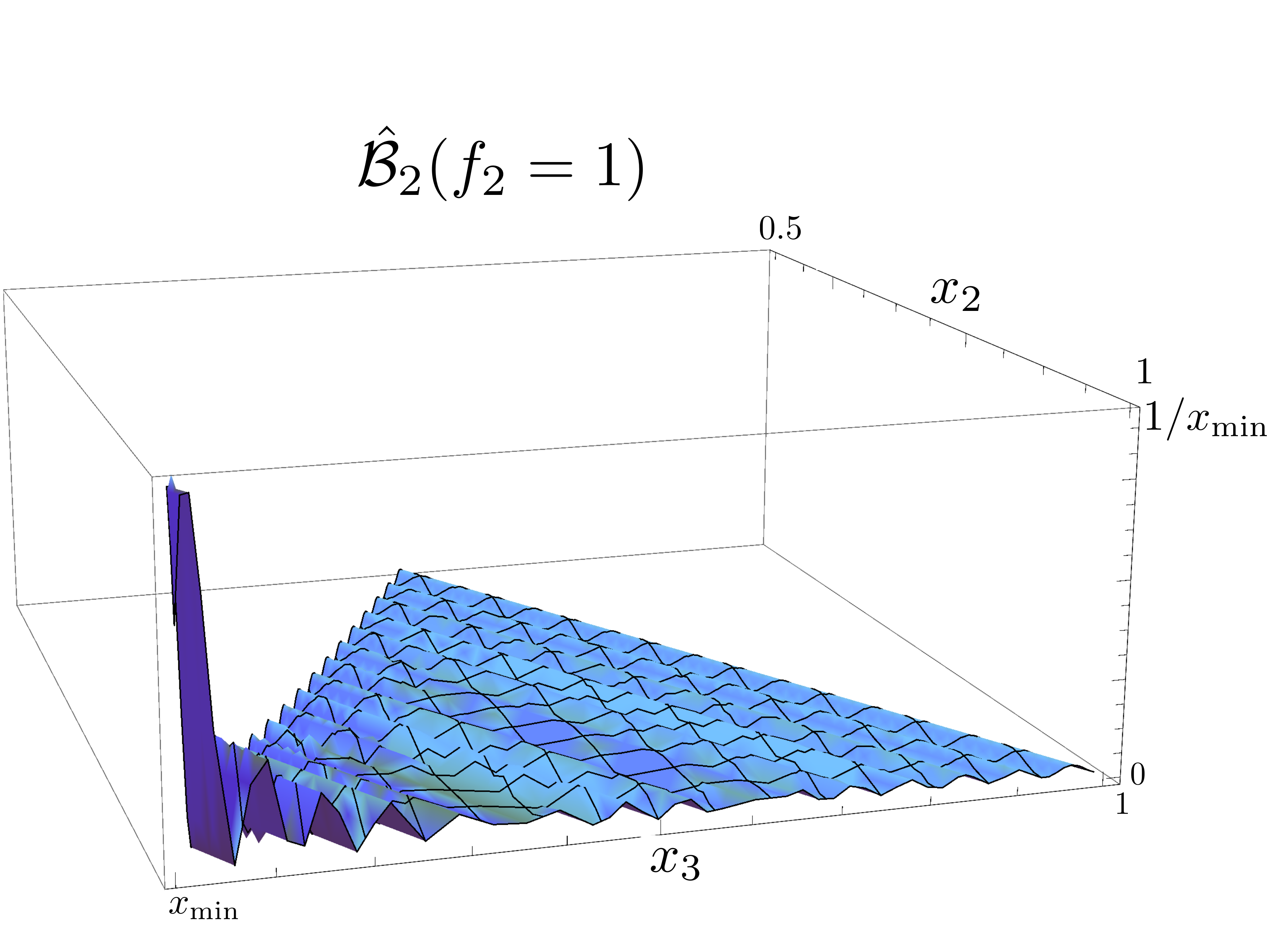} 
\caption{Contribution from the real part of $f_2$ to the  term shown in (\ref{f2term}). This term is constant in lines $\tilde{x}_2={\rm constant}$, and attains its maximum value when $\tilde{x}_2=x_{\rm min}$.
\label{fig:f2}}
\end{center}
\end{figure}

\begin{itemize}

\item {\bf Collinear and nearly collinear configurations:}
The term containing $1/{\tilde{x}_{1}}$ in Eq.(\ref{BGIS}) is  proportional to 
\be \label{f1term}
\hat{\mathcal{B}}_1={\rm Re} \left[ f_1 \, \frac{1-e^{i \,\tilde x_1/x_*}}{\tilde x_1}\right] \, .\ee
This term  is constant along the lines $\tilde{x}_1={\rm const.}$, and produces a large contribution when $\tilde{x}_1\to 0$. Since $\tilde{x}_{1}=-1+x_2+x_3$, this limit corresponds to configurations where the three momenta are collinear. In Figure (\ref{fig:f1f2}) we show the shape of this term. The contribution of this term to the bispectrum in the $\tilde{x}_1\to 0$ limit is proportional to ${\rm Im}(f_1)\frac{1}{x_*}$.
However, notice that the exact limit $\tilde{x}_1\to 0$ is {\it not} the point where the part proportional to ${\rm Re}(f_1)$ in this term takes its maximum amplitude. Instead, the contribution proportional to ${\rm Re}(f_1)$ is largest when $\tilde{x}_1\approx 2.33\;x_*$. This contribution, that corresponds to very nearly, but not exactly, collinear momenta configurations, is proportional to ${\rm Re}(f_1)\frac{0.72}{x_*}$. In summary, the term shown in (\ref{f1term}) attains its largest value along two different lines
 \be
\hat{\mathcal{B}}_{1}\rightarrow\left\{ \begin{array}{ll}
        \frac{1}{x_*}  {\rm Im}(f_1),&\hspace{3mm}\; \tilde{x}=0 \, {\rm  \, (collinear \, line)};  \\
\frac{1}{x_*}[0.72\,{\rm Re}(f_1)+ 0.31\,{\rm Im}(f_1)] & \hspace{3mm}\; \tilde{x}\approx 2.33 \, {\rm  \, (nearly \, collinear \, line)}
  \end{array} \right.
\ee

\item {\bf Isosceles Squeezed Configurations:} The term containing $1/{\tilde{x}_{2}}$ in Eq.(\ref{BGIS}) is proportional to 

\be \label{f2term}
\hat{\mathcal{B}}_2={\rm Re} \left[ f_2 \, \frac{1-e^{i \tilde x_2/x_*}}{\tilde x_2}\right] \, . \ee

This term  is constant along the lines $\tilde{x}_2={\rm const.}$, and produces the largest contribution when $\tilde{x}_2\to 0$. However, since $\tilde x_2=1-x_2+x_3$, the minimum value it  can take is $\tilde x_2=x_{\rm min}$, that corresponds to squeezed configurations where $x_3$ becomes much smaller than the other momenta, $x_3=x_{\rm min}\ll 1\,(=x_1)\approx x_2$. 
 The largest contribution of this term is thus proportional to $1/x_{\rm min}$. Note the exponential is always highly oscillatory because  $x_{\rm min}\gg x_*$, and cannot contribute. Figure (\ref{fig:f2}) shows the part of this term proportional to ${\rm Re}(f_2)$. In summary, the largest value of (\ref{f2term}) is:
\be
\hat{\mathcal{B}}_{2}\rightarrow{\rm Re}(f_2)\frac{1}{x_{\rm min}}\; \hspace{5mm} \; \tilde{x}_2= x_{\rm min} \  {\rm (squeezed \, line)}.
\ee
Note that the part proportional to ${\rm Im}(f_2)$ averages to zero as a result of the highly oscillatory exponential.

\end{itemize}

Considering both contributions, Eq.(\ref{f1term}) and Eq.(\ref{f2term}), together, we find that the bispectrum $B_{\rm GIS}$ takes its largest value in squeezed configurations that are exactly or nearly collinear. For exactly collinear-squeezed configuration we have (expressing the result in terms of the $k$'s)
\ba \label{collinear}
B_{\rm GIS}={\mathcal{B}}_{\rm GIS} \, P_{\zeta}(k_1)P_{\zeta}(k_{\rm min}) \, \left[{\rm Im} (f_1) \, \frac{k_1}{k_*}+{\rm Re}  (f_2) \,  \frac{k_1}{2 \,  k_{\rm min}} \right] \,  ,
 \ea
with $k_1=k_2+k_{min}$ (and $k_3\approx 2 k_{\rm min}\ll k_1, k_2$). And for nearly collinear-squeezed configurations
\ba \label{nearly collinear}
B_{\rm GIS}= {\mathcal{B}}_{\rm GIS} \, P_{\zeta}(k_1)P_{\zeta}(k_{\rm min}) \, \left[\big({\rm Re}  (f_1) \, 0.72 + {\rm Im}  (f_1) \, 0.31\big)  \, \frac{k_1}{k_*}+{\rm Re}  (f_2) \,  \frac{k_1}{2 \,  k_{\rm min}} \right] \,  , 
\ea
with $-k_1+k_2+k_{\rm min}= 2.33 \, k_*$ (and $k_3\approx 2 k_{\rm min}\ll k_1,k_2$). Notice that the momentum-dependent pre-factors $P_{\zeta}(k_1)P_{\zeta}(k_{\rm min})$ significantly enhance squeezed configurations relative to the others, but do not otherwise change the analysis above. Note also that because $k_{\rm min}\gg k_*$, the collinear and nearly collinear-squeezed configurations are much more important than the isosceles-squeezed configuration.

\begin{figure}[h]
\begin{center}
$\begin{array}{cc}
\includegraphics[width=0.5\textwidth,angle=0]{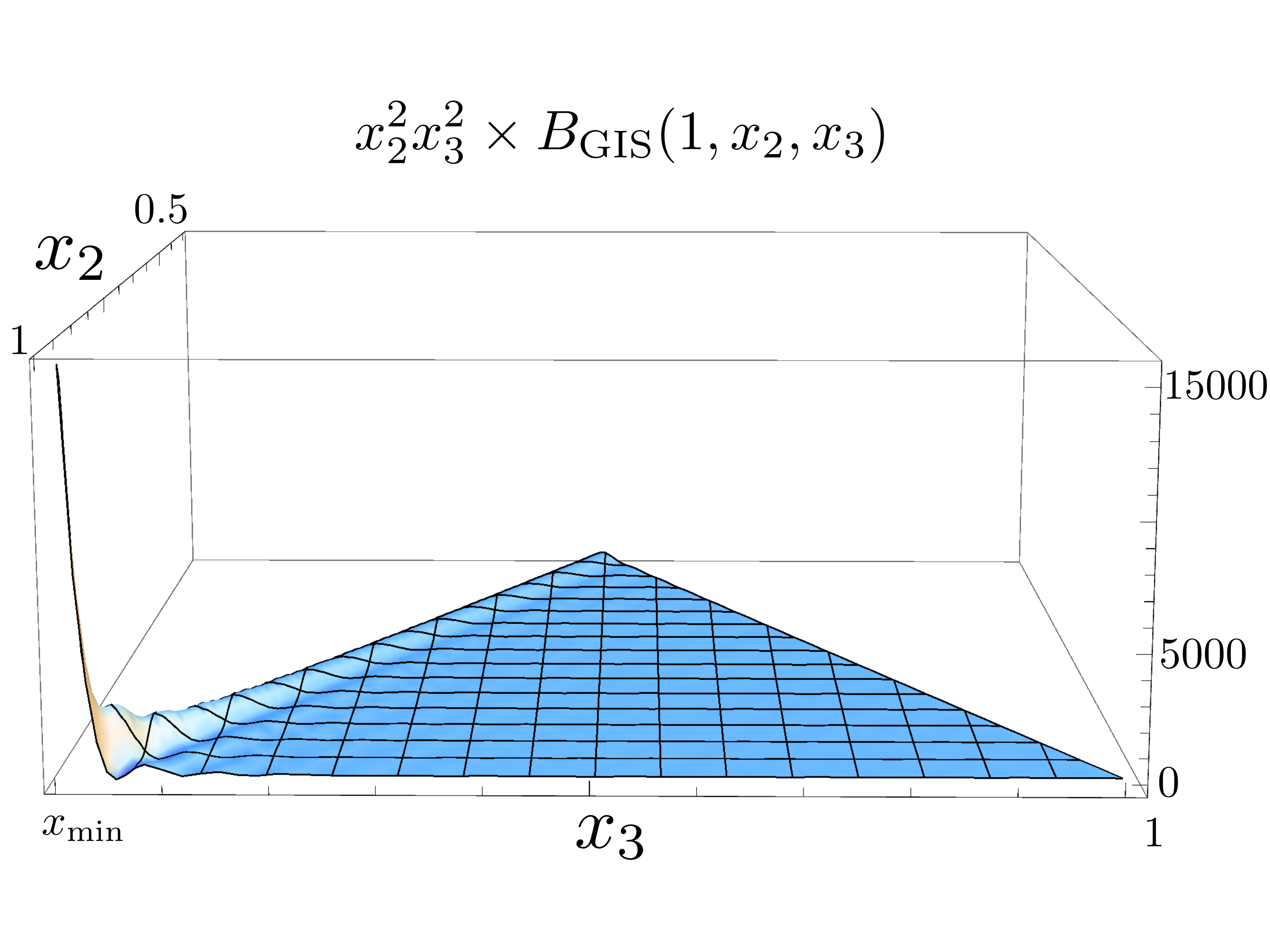} 
&
\includegraphics[width=0.5\textwidth,angle=0]{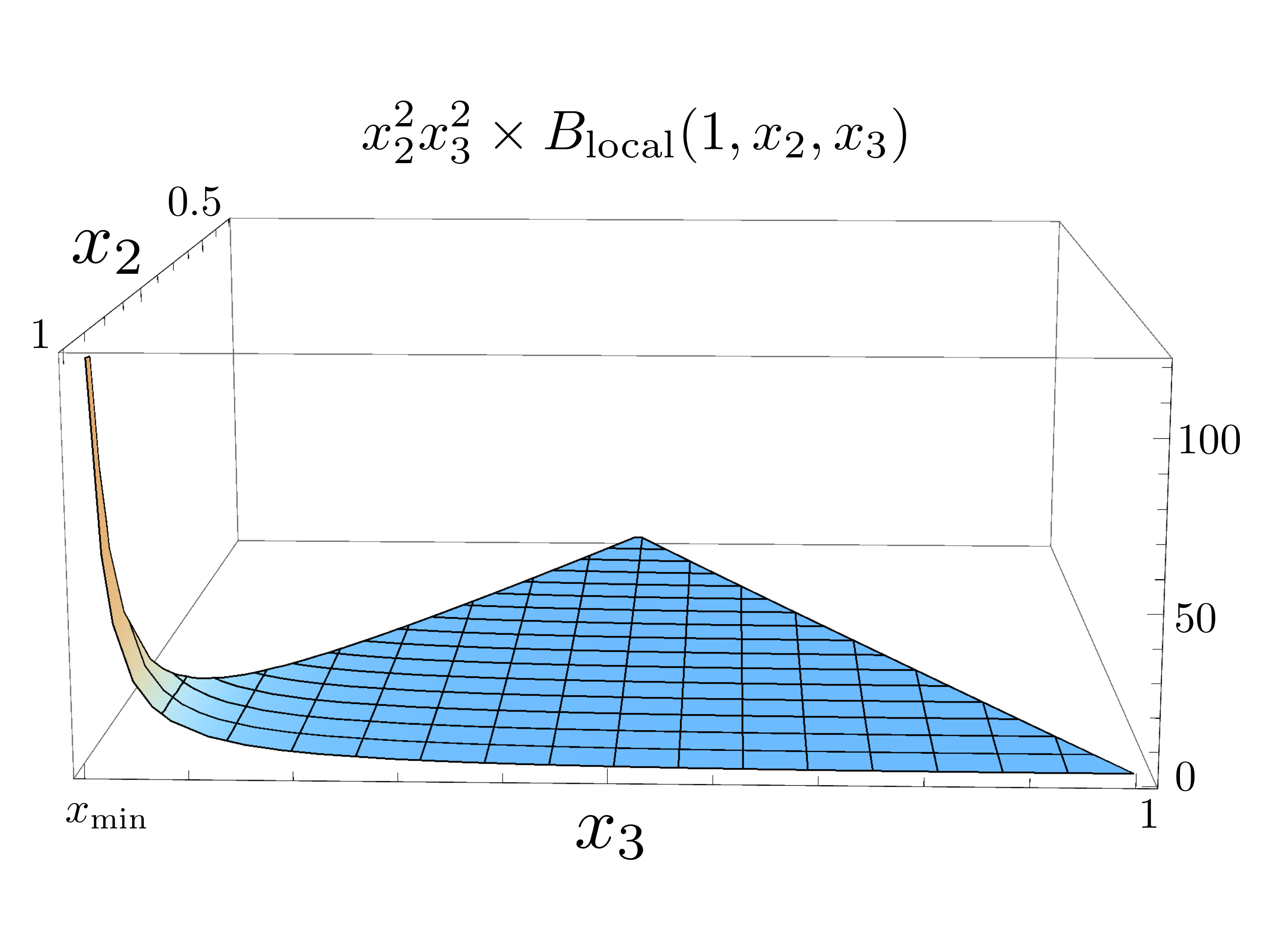} 
\end{array}$
\caption{$B_{\rm GIS}$ (left) with $f_i=1+i$ and $x_*=10^{-2}$, and the local ansatz $B_{\rm local}$ (right). Both bispectra are normalized by ${\cal{B}}_{\rm GIS}={\cal{B}}_{\rm local}=6/5$ (that corresponds to $f_{NL}=1$), and have been multiplied by the factor $x_2^2 x_3^2$. $B_{\rm local}$ is large in squeezed configurations. $B_{\rm GIS}$ is largest in all collinear configurations, with a very significant over-all enhancement in squeezed triangles.\label{fig:BGISandLocal}}
\end{center}
\end{figure}

\begin{figure}[h]
\begin{center}
\includegraphics[width=0.7\textwidth,angle=0]{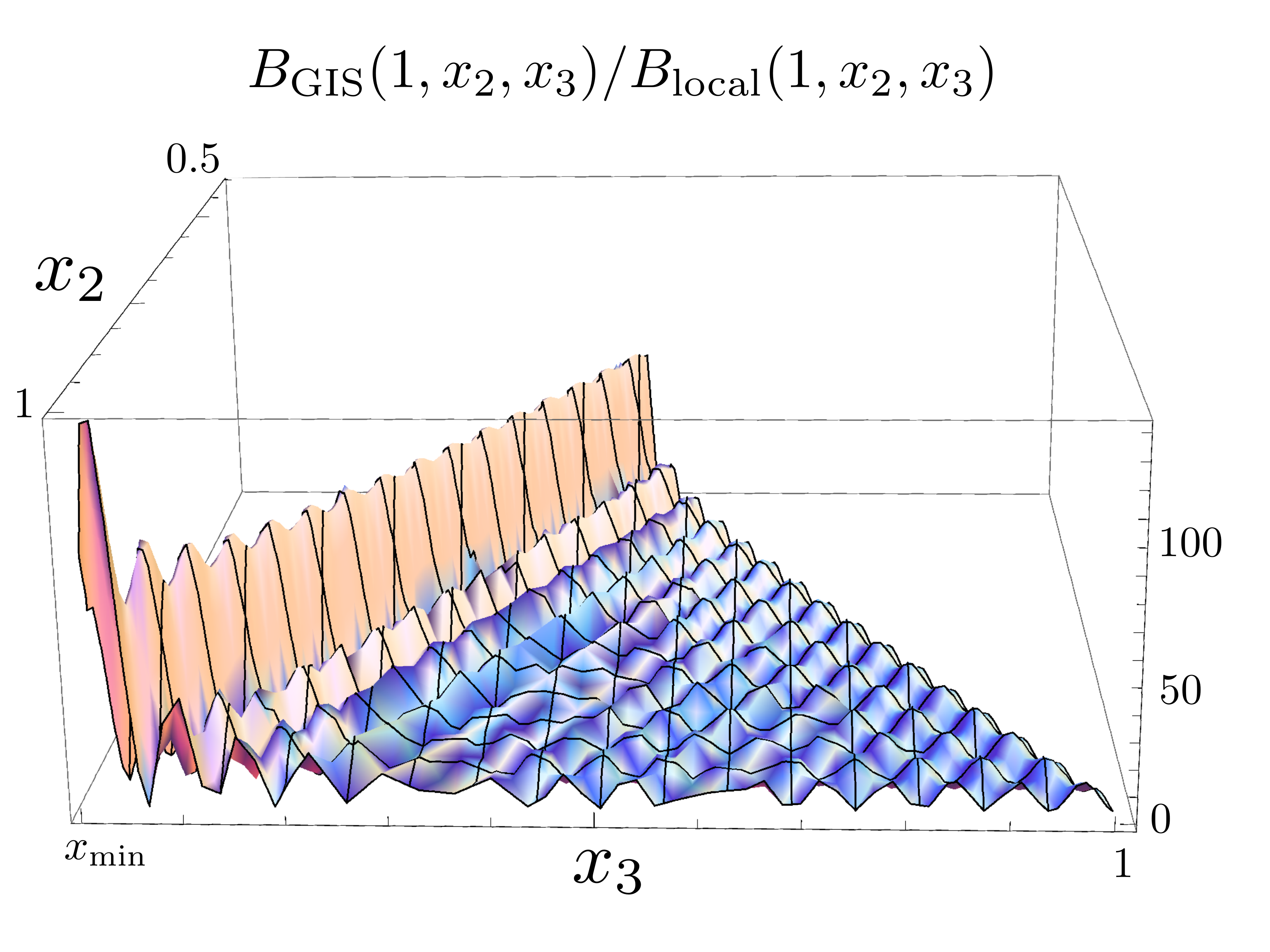} 
\caption{$B_{\rm GIS}/B_{\rm local}$ with $f_i$ real. This figure shows the importance of collinear configurations, with an additional enhancement in the squeezed limit, to $B_{\rm GIS}$ compared to $B_{\rm local}$.
\label{fig:BGISvsLocal}}
\end{center}
\end{figure}

In Figures (\ref{fig:BGISandLocal}) and (\ref{fig:BGISvsLocal}) we plot the total bispectrum $B_{\rm GIS}$ (Eq. (\ref{BGIS})) and compare it with the local ansatz (Eq.(\ref{local})). In those plots we can see how $B_{\rm GIS}$ is characterized by a larger amplitude in all collinear configurations and an over-all enhancement in squeezed triangles. \\

It is worth to emphasize that Eq.(\ref{collinear}) and Eq.(\ref{nearly collinear}) show the largest numerical value that $B_{\rm GIS}$ attains, which correspond to the squeezed configurations that are collinear or very near collinear. However, it is not necessarily true that those are the configuration that most contribute to physical observables such as the non-Gaussian bias (Section \ref{sec:LSS}), which depend on an {\em integral} over some weighted configurations of $B_{\rm GIS}$. 
To better understand what  the most relevant configurations are, it is useful to re-write $B_{\rm GIS}$ as follows. We focus in the squeezed limit,  $k_3\ll k_1,k_2$, because this is the most significant region. We can use the triangle condition, $\vec{k}_1+\vec{k}_2+\vec{k}_3=0$, to write $k_2=\sqrt{k_1^2+k_3^2+2k_1k_3\mu}$, where $\mu={\rm cos}\,(\theta_{13})$ (the angle between $\vec{k}_1$ and $\vec{k}_3$). The bispectrum in the squeezed limit  can then be written as

\ba \label{mu}
B_{\rm GIS}(k_1,k_2,k_3)\rightarrow{\mathcal{B}}_{\rm GIS} \, P_{\zeta}(k_1)P_{\zeta}(k_3)\,\frac{k_1}{k_3} {\rm Re} \left[f_1 \frac{1-e^{i(1+\mu)k_3/k_{*}}}{1+\mu} +f_2\frac{1-e^{i(1-\mu)k_3/k_{*}}}{1-\mu}\right] \nonumber \, .\ea
This expression shows again that the maximum numerical value for the bispectrum corresponds to collinear configurations ($\mu\to\pm1$) that are squeezed ($k_3\to k_{\rm min}$), and the formulas (\ref{collinear}) and (\ref{nearly collinear}) can be easily obtained from it in that limit. Note also the global $1/k_3$ dependence. Additionally, the above expression shows that the configurations with larger ``volume", i.e. most contributing to the integral of  $B_{\rm GIS}$, are those for which $|\mu|$ is close to 1, so the denominators in (\ref{mu}) are small, but still small enough so the oscillatory behavior of the exponentials is important. Those configurations are squeezed but between nearly collinear and isosceles triangles (sometimes called `elongated' in the literature) and will provide the largest contributions to observables such as the halo bias.

\subsection{Examples and constraints for the initial state}
\label{sec:examples}

In a scenario where inflation starts at some finite time, the state of perturbations at the onset of inflation may deviate from the vacuum state as a consequence of a non-trivial pre-inflationary evolution. There are several examples in the literature of initial states obtained, for instance, by assuming a radiation dominated phase before inflation \cite{Vilenkin:1982wt}, an anisotropic pre-inflationary phase of the universe \cite{Dey:2011mj}, the (singularity-free) pre-inflationary spacetime arising from Loop Quantum Cosmology \cite{AAN}, initial states arising from integrating out heavy fields \cite{Baumann:2011su,Shiu:2011qw,Achucarro:2012sm, Jackson:2012fu,Avgoustidis:2012yc,Cespedes:2012hu}, etc. Generically, pre-inflationary evolution could produce a state for 
the perturbations with some number of quanta compared to the Bunch-Davies state, and which need not be Gaussian or pure. Therefore, the most general state would be a non-vacuum, non-Gaussian, mixed quantum state. Because our goal is to study the non-Gaussianity generated during inflation, for simplicity we consider a Gaussian, pure initial states (see \cite{Agullo:2010ws} for a discussion including mixed states.) These states can be described by Bogoliubov transformations of the vacuum.

To specify the initial state we first expand the field operator of the primordial curvature perturbation $\zeta$ in Fourier modes
\be {\zeta}(\vec{x},\tau)=\int \frac{d^3 k}{(2\pi)^3} \ { \zeta}_{\vec{k}}(\vec{x},\tau) \ ,   \hspace{1cm}  {\zeta}_{\vec{k}}(\vec{x},\tau)=(A_{\vec{k}} \, \bar{\zeta}_{k}(\tau)+ A_{-\vec{k}}^{\dagger} \, \bar{\zeta}_{k}^*(\tau)) \ e^{i \vec{k}\vec{x}}\ ,\ee 
where $\tau$ is conformal time. The mode functions can be written as a linear combination
$\bar{\zeta}_{k}(\tau)=\alpha_k \ {\zeta}_{k}(\tau)+\beta_k \ {\zeta}_{k}^*(\tau)$, where $|\alpha_k|^2-|\beta_k|^2=1$, 
and
\be \label{modes}{\zeta}_k(\tau)= \frac{H^2}{\dot \phi_0} \frac{(1+ik \tau)}{\sqrt{2 k^3}}e^{-ik\tau} \ , \ee
are de Sitter invariant modes.\footnote{In an inflationary spacetime that departs from the pure de Sitter geometry the above modes generalize to Hankel functions (see, for instance, \cite{Weinberg:2008zzc}).} Here $\phi_0(t)$ is the homogeneous part of the inflaton field. The Hubble scale is $H\equiv\dot a/a$ with $a(t)$ the scale factor and the dot indicates the derivative with respect to the cosmic time. With the normalization (\ref{modes}), the operators $A_{\vec{k}}$ and $A_{\vec{k}}^\dagger$ satisfy
commutation
relations $[A_{\vec{k}},A_{\vec{k'}}^\dagger]=(2\pi)^3\delta^3(\vec{k}-\vec{k'})$. When $\alpha_k=1$ and $\beta_k=0$ for all $k$, the state annihilated by all the operators $A_{\vec{k}}$ is called the Bunch-Davies vacuum state. For arbitrary values of the Bogoliubov coefficients $\alpha_k$ and $\beta_k$ 
(with $|\alpha_k|^2-|\beta_k|^2=1$) the state is a Bogoliubov transformation of the Bunch-Davies vacuum. It can be interpreted as containing an average number density of quanta per unit proper volume
$(2\pi a)^{-3}|\beta_k|^2d^3k$ with momenta near $\vec{k}$ in the range $d^3k$, as compared to the Bunch-Davies vacuum.

There are restrictions on behavior of the coefficients $\alpha_k$ and $\beta_k$ coming both from theoretical considerations and CMB observations. To show this  explicitly, we consider the representative example in which the number of initial quanta fall off with some power of $k$

\be \label{profile} N_k\equiv |\beta_k|^2=N_0 \left(\frac{k}{k_*}\right)^{-\hat\delta} \, , \ee
where $k_*\equiv a(\tau_0) H$ (any other choice for this scale translates into a redefinition of the constant $N_0$).

\begin{itemize} 

\item{\bf Renormalizability condition}

From the theoretical point of view, the adiabatic condition \cite{Parker:Toms}, or similarly the Hadamard condition, restricts the ultra-violet behavior of the initial state by requiring that $\alpha_k \to 1$ and $\beta_k \to 0$ faster than $k^{-4}$, when $k \to \infty$. This condition ensures that the UV divergences appearing in expectation values of relevant quantum operators can be systematically cured by methods of renormalization and regularization. However, it does not necessarily restrict $N_k$ for the finite values of $k$ relevant for observations. We consider then

\be 
\hat\delta= 
\left\{ \begin{array}{ll}
         \delta& {\rm for}\ k<k_{\rm max}\\
    \delta_0> 4 &  {\rm for}\ k \ge k_{\rm max} \; \end{array} \right.
\ee
for some scale $k_{\rm max}>k_*$ and, for simplicity, we consider both $\delta$ and $\delta_0$ constants. Notice that since physically reasonable initial states can only deviate from Bunch-Davies over a finite range of  $k$, a modified initial state will not alter the usual consistency relation in the limit $k_3/k_1\rightarrow0$. 

\item{\bf Negligible backreaction condition}

The initial state must satisfy that the back-reaction of its energy density should not modify the inflationary background expansion. The expression for the energy density of the generalized initial state can be obtained by considering the time-time component of the renormalized stress-energy tensor (see, for instance, \cite{Bunch:1978yq, Anderson:1987yt, Anderson:2005hi} for explicit expressions). One can, however, obtain a reasonable estimate by using 

\be
\rho_{\rm N(k)}(\tau)\approx \frac{1}{a(\tau)^4}\int_0^{\infty} d^3k\;k\, N_k\;.
\ee
We will demand that at any time $\tau$ during inflation $\rho_{\rm N(k)}(\tau)$ has to be negligible compared to the energy density of the unperturbed part of the inflation field, which is given by $\rho_0\approx M_P^2 H^2$.
A stronger condition, however, is obtained by demanding that the change in time of the energy density $\rho_{\rm GIS}$ to be small, in such a way that the slow-roll conditions are not violated. This gives \cite{Greene:2004np,Holman:2007na}

\be \frac{\rho_{\rm N(k)}(\tau)}{\rho_0}\lsim \epsilon= {\mathcal{O}}(10^{-2})\, . \ee

From this inequality the following restrictions for $N_0$ are obtained (we show here a few examples)
\begin{itemize}
\item For $\delta=\delta_0>4$, $N_0\lsim \frac{\epsilon M_P^2}{H^2} (\delta_0-4)\approx 10^9 \, (\delta-4)$.
 \item For $\delta=2$, $N_0\lsim \frac{\epsilon M_P^2}{H^2} \left(\frac{k_*}{k_{\rm max}}\right)^{2} \approx 10^5 $,
 \item For $\delta=0$, $N_0\lsim \frac{\epsilon M_P^2}{H^2} \left(\frac{k_*}{k_{\rm max}}\right)^{4} \approx 10 $,

\end{itemize}
 
 where we have used $k_{\rm max}\approx 10^2 \, k_*$. 
 
 Note that negligible back reaction and the renormalizability condition imply small number of all initial quanta for observable modes if inflation lasts much longer than the standard assumption  of around 65  e-folds. 

\item{\bf Spectral index condition}

The observation of a nearly scale invariant power spectrum  in the CMB imposes the strongest condition on $N_0$. The power spectrum arising from a generalized initial state is given by
\ba P_{\zeta}(k)=|\bar{\zeta}_k|^2=\frac{1}{2 \epsilon M_P^2} \frac{H^2}{2 k^3} |\alpha_k+\beta_k|^2\ .\ea
The spectral index is then

\be \label{index} n^{GIS}_s-1\equiv \frac{d\ln{(k^3 P_{\zeta}(k))}}{d\ln{k}}= n^{(0)}_s -1
+ \frac{d\ln{|\alpha_k+\beta_k|^2 }}{d \ln k} \ , \ee

where $n^{(0)}_s$ is the spectral index obtained in the vacuum state computation. The observed \cite{Komatsu:2010fb} value $n_s=0.963\pm 0.012$, imposes the condition
$$\left| \frac{d\ln{|\alpha_k+\beta_k|^2}}{d \ln k}\right| \lsim {\mathcal{O}}(10^{-2})\, .$$
This inequality, for $k\approx k_*$, requires

\be 2 N_0 (\delta-10^{-2})\lsim 10^{-2}\, .\ee
Note that this condition severely restricts the size of $N_0$, unless the value of $\delta$ is close to $10^{-2}$.
\end{itemize}

It is interesting to estimate the way the squeezed limit of the bispectrum scales with momenta for the example in Eq.(\ref{profile}). To simplify the equations, we will consider $\mu\approx0$ and $k_3\ll k_1\approx k_2$. As we saw at the end of section (\ref{sub:shape}), this point is not where the amplitude of the bispectrum is maximum, but it is representative of the dominant behavior of the bispectrum, especially for Large Scale Structure observables. In that case
\ba
B(k_1,k_2,k_3)&\rightarrow&\\ & &\hspace{-2cm}{\mathcal{B}}_{\rm GIS} \, P_{\zeta}(k_1)P_{\zeta}(k_3) \, {\rm Re}\left[ \frac{f_t+f_3}{2} +\frac{k_1}{k_3}f_1\left(\frac{1-e^{i(1+\mu)k_3/k_{*}}}{1+\mu}\right) +\frac{k_1}{k_3}f_2\left(\frac{1-e^{i(1-\mu)k_3/k_{*}}}{1-\mu}\right)\right]\nonumber \\\nonumber 
&&\hspace{-2cm}\approx {\mathcal{B}}_{\rm GIS} \, P_{\zeta}(k_1)P_{\zeta}(k_3)\left[ \frac{f_t+f_3}{2} +\frac{k_1}{k_3}(f_1+f_2)\right]\;,{|\mu|\approx 0}\;.
\ea
To make the discussion more transparent, let us consider $|\alpha_k+\beta_k|^2\approx 1+|\beta_k|^2$ (i.e. we neglect the interference terms between $\alpha$ and $\beta_k$).
This, combined with the fact that in the  squeezed limit $k_1\approx k_2$, allow us to simplify the expressions for the $f_i$, and we find 
\ba 
f_t+f_3&=&1\, , \\ \nonumber
f_1+f_2&\approx&2f_1= \frac{2N_{k_1}(1+N_{k_1})}{(1+2N_{k_1})(1+2N_{k_3})} \, .
\ea
If the number of initial quanta is small, $N_k\ll1$, the dominant behavior of the bispectrum is captured by
\be
\label{eq:BwithLargeN}
B(k_1,k_2,k_3)\sim {\mathcal{B}}_{\rm GIS} \, P(k_1)P(k_3)\left[1+2N_0 \frac{k_1}{k_3} \left(\frac{k_1}{k_*}\right)^{-\delta}\right] \, .
\ee
In case $N_k> 1$ we have, instead

\ba
B(k_1,k_2,k_3)&\sim&{\mathcal{B}}_{\rm GIS} \, \, P_{\zeta}(k_1)P_{\zeta}(k_3) \frac{1}{2}\frac{k_1}{k_3} \left(\frac{k_3}{k_1}\right)^{\delta} \, ,\\\nonumber
\ea
where  in this case $\delta\lsim 10^{-2}$ as a consequence of the spectral index constraint. However, we see that for $\delta>0$, the bispectrum is {\it at most} as divergent as $1/k_3^4$ in total.

\section{Observational Consequences for Large Scale Structure}
\label{sec:LSS}
Although several authors have looked at how the CMB might constrain the GIS bispectrum \cite{Holman:2007na,Meerburg:2009ys,Meerburg:2009fi,Meerburg:2010rp,Ganc:2011dy}, the strong enhancement in the squeezed limit, Eq.(\ref{mu}) above, means that Large Scale Structure should already provide an excellent constraint. In this section we compute the LSS signatures, beginning with the effect on the power spectrum of dark matter halos and galaxies (the non-Gaussian bias). We also compute the total skewness, which gives a feeling for how non-Gaussian the GIS scenario is and how much the expected number density of galaxies and galaxy clusters is affected. In this section, we normalize the amplitude of both bispectra by ${\cal{B}}_{\rm GIS}={\cal{B}}_{\rm local}=\frac{6}{5}$ (corresponding to $f_{NL}=1$) and we write $f_i=1+i$ to simplify the parameter space.
\subsection{Cosmology}
We use WMAP 7 year best fit values for parameters of the homogeneous cosmology (including the matter density $\Omega_m$ and the Hubble parameter today, $H_0$) and the fluctuations \cite{Komatsu:2010fb}. On large scales the power spectrum is well described by $\mathcal{P}_{\zeta}(k)\approx2.42\times10^{-9}\left(\frac{k}{0.002{\rm Mpc}^{-1}}\right)^{n_s-1}$, with the spectral index $n_s-1\approx - 0.034$. 

The relation between the primordial curvature perturbation $\zeta$ and the linear perturbation to the matter density $\delta =\delta \rho/\rho$ today is
\ba
\delta(\vec{k},z)&=&\frac{3}{5}M(k,z)\zeta(\vec{k})=M(k,z)\Phi(\vec{k}) \, ,\ea
with \ba M(k,z)&=&\frac{2}{3}\frac{1}{\Omega_{m}}\frac{1}{H_0^2} \, D(z)\, T(k)\, k^2
\label{Mka}
\, , \ea
where $\Phi(k)$ is the Bardeen potential, $D(z)$ is the linear growth function, $z$ is the redshift, and $T(k)$ is the transfer function. The smoothed density field is given by
\be
\delta_R(z)=\int \frac{d^3k}{(2\pi)^3} W_R(k)\delta(\vec{k},z) \, ,
\ee
where $W_R(k)$ is the Fourier transform of a window function.  
Since we compute the statistics of the smoothed density field, it is useful to define the $M_R(k,z)=M(k,z)W_R(k)$. The smoothed variance is then
\ba
\label{eq:smoothsig}
\sigma^2_R=\langle\delta^2_R\rangle&=&\int\frac{d^3k}{(2\pi)^3}\int\frac{d^3{k^{\prime}}}{(2\pi)^3} \,\frac{9}{25}M_R(k)M_R(k')\langle\zeta_{\vec{k}}\zeta_{\vec{k}'}\rangle\nn
&=&\int_0^\infty \frac{dk}{k}\,\frac{9}{25}M_R(k)^2 \mathcal{P}_\zeta(k)\;.
\ea

\subsection{Generalized initial states and halo bias}
\label{sec:bias}
Large Scale Structure surveys measure the statistics of gravitationally bound objects observed in the late universe. Even if the primordial density perturbations were Gaussian, the power spectrum of the bound objects, formed from sufficiently overdense regions, is not identical to the power spectrum of the linear density field. The ratio of clustering of objects to that of the underlying density field is characterized by the bias, $b$ (see \cite{Bernardeau:2001qr} for a comprehensive review). For example, the power spectrum of dark matter halos, $P_{hh}$, can be related to the matter power spectrum $P_{mm}$ by
\be
P_{hh}(k)=b^2P_{mm}(k) \, ,
\ee
where on large scales $b$ is roughly scale-independent but depends on the mass of the halo. 

The bias seen on large scales can shift significantly if the primordial perturbations are non-Gaussian. Although anticipated by early theoretical work \cite{Grinstein:1986en, Matarrese:1986et}, this was first definitively seen in simulations of the exact local ansatz \cite{Dalal:2007cu}, where a strongly scale-dependent term in the bias was uncovered. Further analytic and simulation work \cite{Matarrese:2008nc, Desjacques:2008vf, Pillepich:2008ka, Giannantonio:2009ak} verified that the leading new contribution to the bias on large scales (small $k$) from local type non-Gaussianity is
\ba
\Delta b_{\rm NG, local}(k,z)&=&\frac{2f_{NL}b_1\delta_c}{M(k,z)}\propto\frac{f_{NL}}{k^2}
\ea
where $b_1$ is the linear Gaussian (Lagrangian) bias, $\delta_c$ is the collapse threshold (1.686 in spherical collapse) and we have used $T(k)\rightarrow1$, so $M(k)\propto\frac{1}{k^2}$ on large scales. Measurements of the power spectrum of objects like quasars and luminous red galaxies have already been used to constrain non-Gaussianity of the local type \cite{Slosar:2008hx, Xia:2010pe} at a level competitive with WMAP, and results from future LSS surveys are expected to match or exceed even the best constraint from Planck satellite measurements of the CMB \cite{Carbone:2008iz, Desjacques:2010jw, Cunha:2010zz}. 

The computation of $\Delta b_{\rm NG}$ from primordial bispectra more general than the local ansatz requires more attention. Since it is the coupling of long and short wavelength modes that leads to the non-Gaussian bias, a good estimation of the signature of {\it any} primordial bispectrum can be obtained by looking at its amplitude in squeezed momenta configurations. Furthermore, simulations \cite{Desjacques:2009jb, Wagner:2010me, Smith:2010gx, Shandera:2010ei} and analytic work \cite{Schmidt:2010gw, Desjacques:2011mq,Desjacques:2011jb} both indicate that the contribution to the bias from the connected primordial $N$-point function can be calculated from
\be
\label{eq:Deltabgeneric}
\Delta b_{NG}(k,z)=\frac{4}{(N-1)!}\frac{\mathcal{F}^{(N)}(k)}{M(k,z)}\left[b_{N-2}\delta_c+b_{N-3}\left(3-N+\frac{d{\rm ln}\mathcal{F}^{(N)}(k)}{d{\rm ln}\sigma_{Rs}}\right)\right]
\ee
for $N\geq3$. The $b_N$ are higher order Gaussian (Lagrangian) bias parameters ($b_0=1$, while the rest of the $b_i$ are numbers determined from simulations or data). The subscript $s$ indicates quantities that are defined locally, on length scales small compared to $k^{-1}$, and the functions $\mathcal{F}^{(N)}$ are related to the $N$-point correlation functions of the Bardeen potential, $\xi_{\Phi}^{(N)}$ by
\ba
\mathcal{F}^{(N)}(k)=\frac{1}{4\sigma_{R_s}^2P_{\Phi}(k)}\left[\prod_{i=1}^{N-2}\int\frac{d^3q_i}{(2\pi)^3}M_{R_s}(q_i)\right]  M_{R_s}(\hat{q})\xi_{\Phi}^{(N)}({\bf q}_1,\dots, {\bf q}_{N-2}, {\bf \hat{q}}, {\bf k}) \, , \ea 
where
\ba {\bf \hat{q}}&\equiv&-{\bf q}_1-\dots-{\bf q}_{N-2}-{\bf k} \, .
\ea

For the contribution from the bispectrum, for example, we have
\ba
\Delta b_{NG}(k,z)&=&2\frac{\mathcal{F}^{(3)}(k)}{M(k,z)}\left[b_1 \, \delta_c+\left(\frac{d{\rm ln}\mathcal{F}^{(3)}(k)}{d{\rm ln}\sigma_{R,s}}\right)\right]+\dots\\\nonumber
\mathcal{F}^{(3)}(k)&=&\frac{1}{4\sigma_{R_s}^2P_{\Phi}(k)}\int\frac{d^3q}{(2\pi)^3}M_{R_s}(q)M_{R_s}(|-{\bf q}-{\bf k}|)B_{\Phi}({\bf q},-{\bf q}-{\bf k}, {\bf k})\\\nonumber
&=&\frac{1}{4\sigma_{R_s}^2P_{\zeta}(k)}\int\frac{d^3q}{(2\pi)^3}M_{R_s}(q)M_{R_s}(|-{\bf q}-{\bf k}|)\frac{3}{5}B_{\zeta}({\bf q},-{\bf q}-{\bf k}, {\bf k})
\ea
Notice that for the local ansatz
\be
\mathcal{F}_{\rm local}^{(3)}(k)\rightarrow f_{NL}
\ee
on large scales (roughly $k\lesssim\; 0.05 h {\rm Mpc}^{-1}$). This is a constant, and produces $\Delta b_{NG, \rm local}\propto \frac{2 b_1f_{NL}\delta_c}{k^2}$.

\begin{figure}[h]
\begin{center}
$\begin{array}{cc}
\includegraphics[width=0.5\textwidth,angle=0]{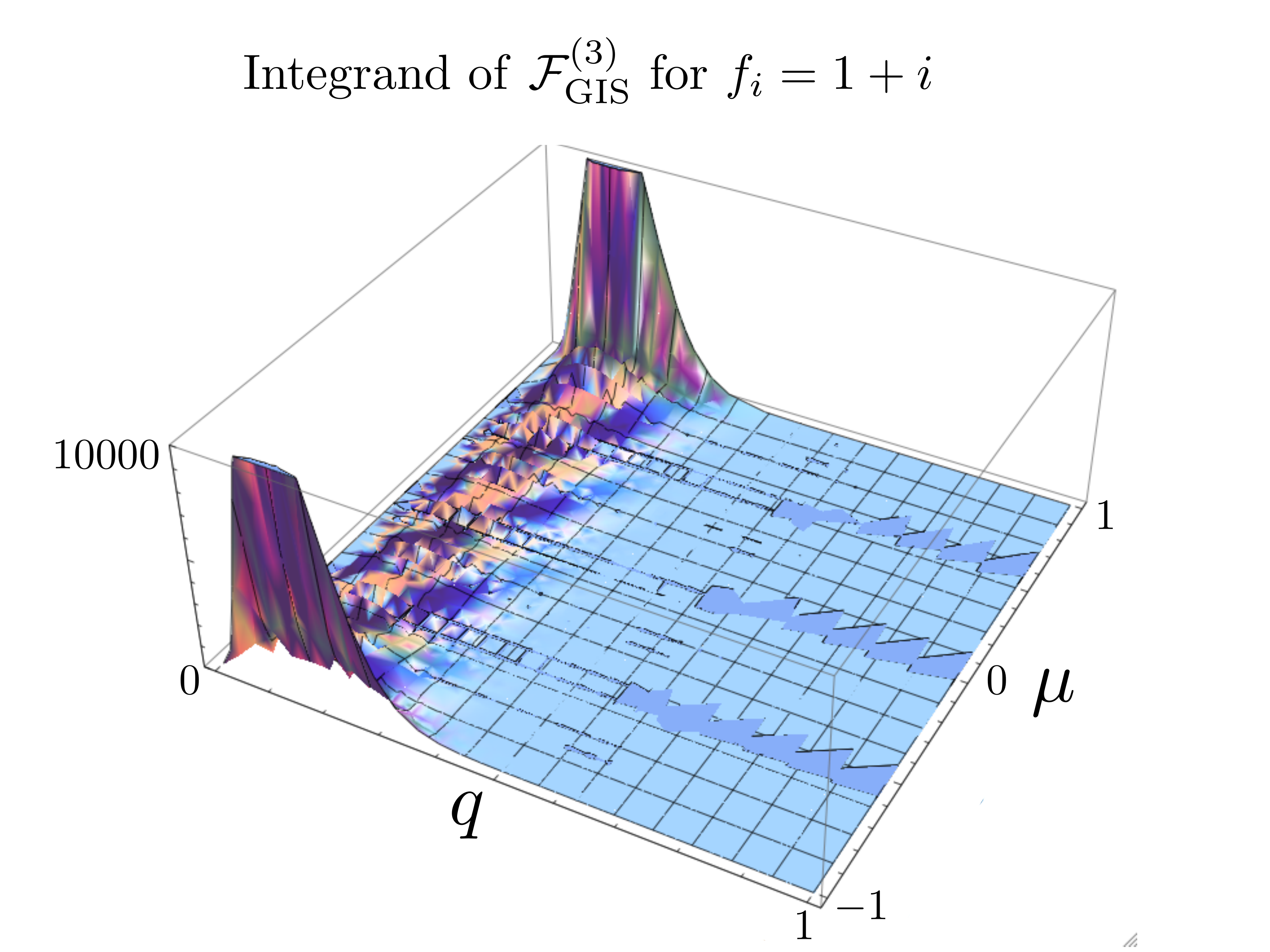} 
&
\includegraphics[width=0.5\textwidth,angle=0]{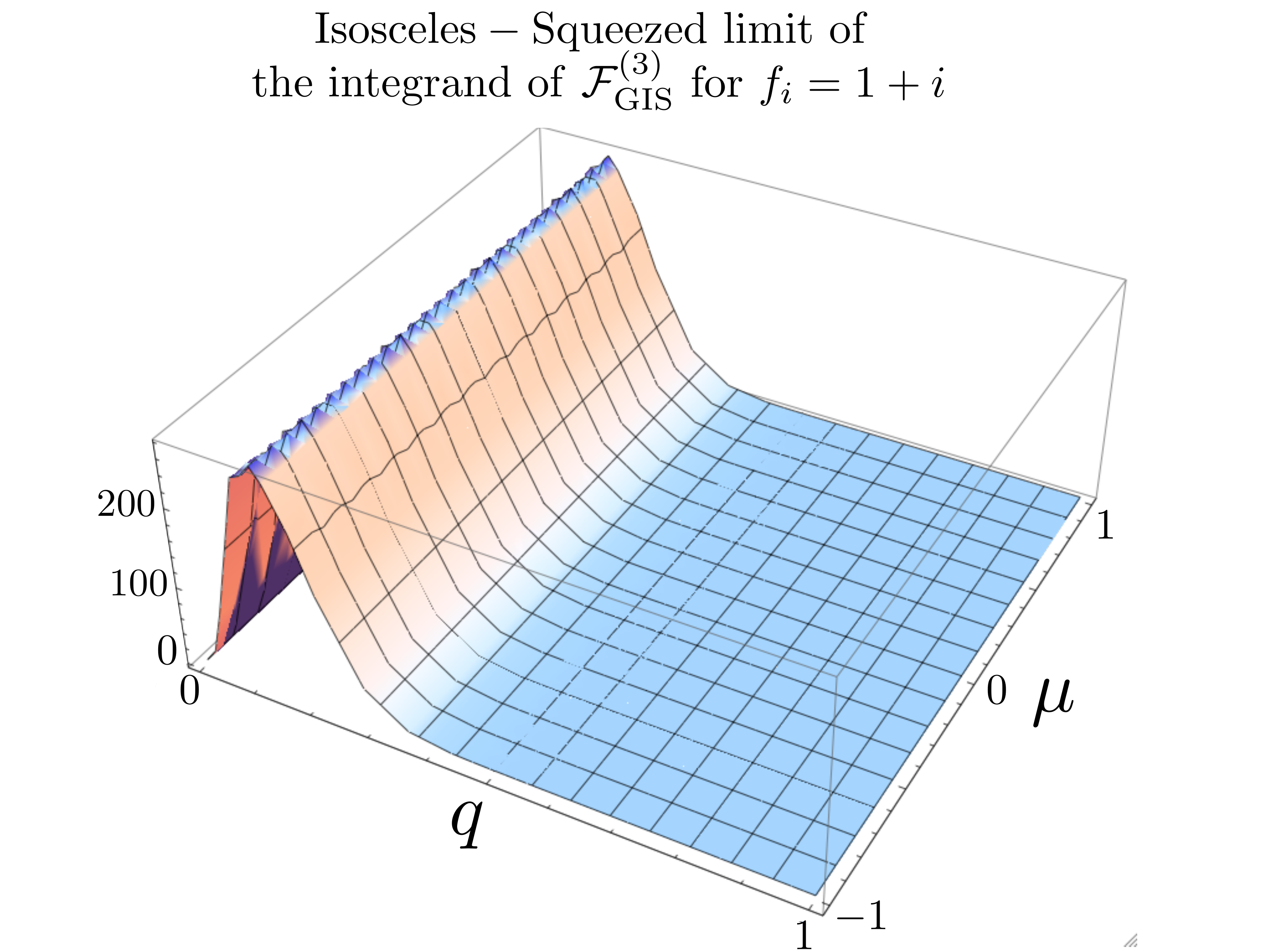} 
\end{array}$
\caption{The left panel shows the full integrand contributing to $\mathcal{F}^{(3)}$ for the Generalized Initial State shape, with $\mathcal{B}_{\rm GIS}=\frac{6}{5}$, $f_{i}=1+i$. The  right panel shows the contribution from the isosceles squeezed limit of the shape only. The significant contributions from the nearly collinear squeezed configurations can be seen from the enhancements near the corners $\mu\approx\pm1$.}
\label{fig:CurlyFcontributions}
\end{center}
\end{figure}

\begin{figure}[h]
\begin{center}
$\begin{array}{cc}
\includegraphics[width=0.5\textwidth,angle=0]{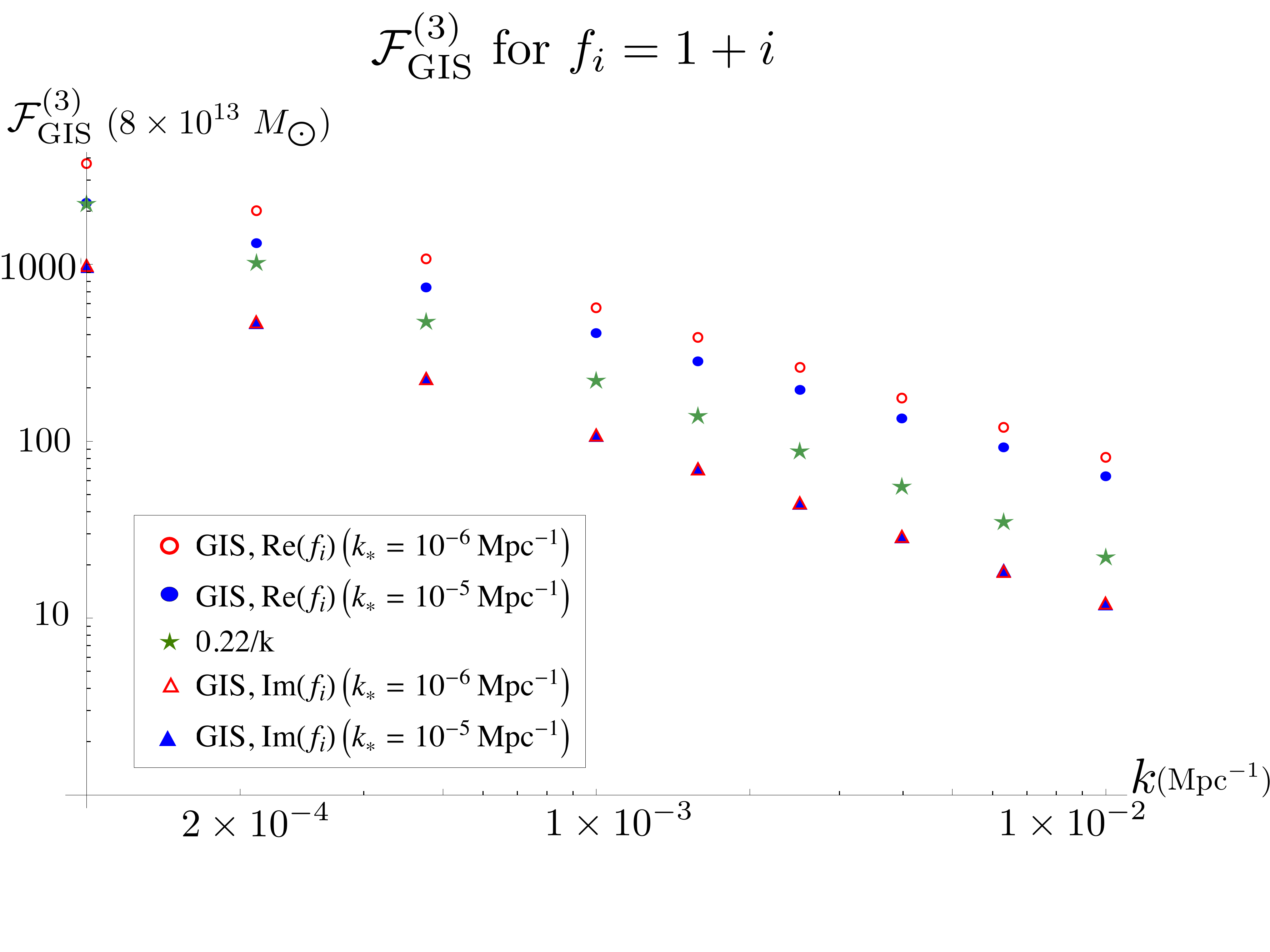} 
&
\includegraphics[width=0.5\textwidth,angle=0]{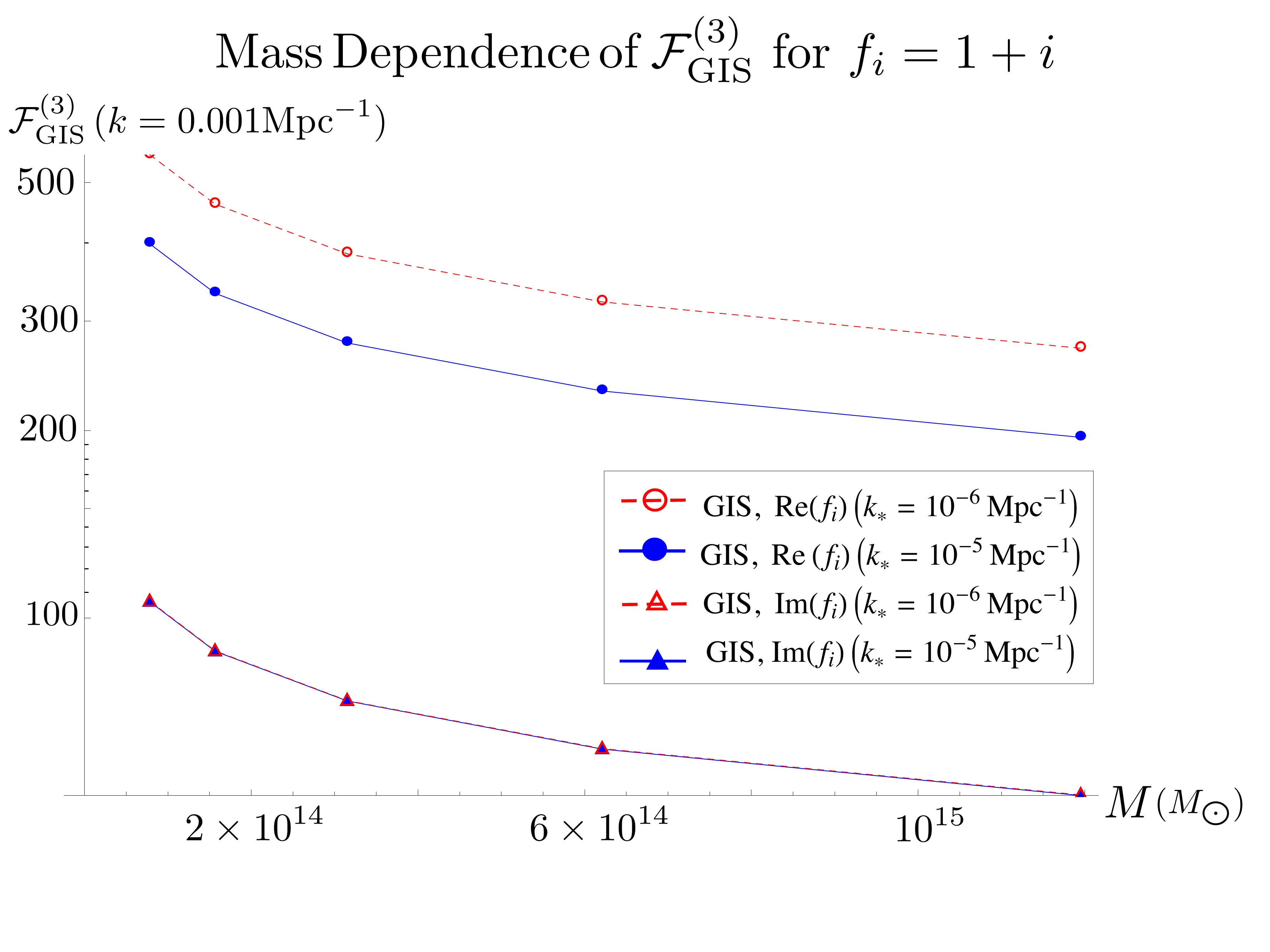} 
\end{array}$
\caption{Left panel: The integral $\mathcal{F}^{(3)}_{\rm GIS}(k, M=7\times10^{13}M_{\odot})$. Right panel: The integral $\mathcal{F}^{(3)}_{\rm GSI}(k=0.001{\rm Mpc}^{-1}, M)$. Both panels show the GIS shape with $\mathcal{B}_{\rm GIS}=\frac{6}{5}$ and $f_i=1+i$. The local ansatz with the same normalization would give a straight line in both panels at $\mathcal{F}=1$. The green stars in the left panel compare the result $\mathcal{F}^{(3)}_{\rm GIS}$ to the roughly expected $1/k$ behavior, normalized to match the $k_*=10^{-5}$ example at $k=10^{-4}{\rm Mpc}^{-1}$. The contributions from real part (circles) and imaginary part (triangles) of $f_i$ are shown, each at two values for the scale $k_*$ (filled blue vs open red points). The Re($f_i$) contribution increases in magnitude as $k_*$ decreases, while the Im($f_i$) contribution stays constant.}
\label{fig:FofM}
\end{center}
\end{figure}

For the Generalized Initial State with constant $f_i$, the expression is
\ba
\label{eq:CurlyFMIS}
\mathcal{F}^{(3)}_{GIS}(k)&=&\frac{3}{5}\frac{ {\mathcal{B}}_{\rm GIS} \, }{16\pi^2\sigma_{R_s}^2}\int_{q_{\rm min}}^{\infty}dq\int_{\mu_{min}}^1d\mu \; q^2M_{R_s}(q)M_{R_s}(\hat{q})\\\nonumber
&&\times\left[P_{\zeta}(\hat{q})\frac{k^2\hat{q}^2}{q^3}+P_{\zeta}(q)\frac{k^2q^2}{\hat{q}^3}+\frac{P_{\zeta}(q)P_{\zeta}(\hat{q})}{P_{\zeta}(k)}\frac{q^2\hat{q}^2}{k^3}\right]\\\nonumber
&&\times\textrm{Re}\left[f_t\frac{1-e^{i(q+k+\hat{q})/k_*}}{q+k+\hat{q}}\right.\\\nonumber
&&\left.+f_1\frac{1-e^{i(-q+k+\hat{q})/k_*}}{-q+k+\hat{q}}+f_2\frac{1-e^{i(q+k-\hat{q})/k_*}}{q+k-\hat{q}}+f_3\frac{1-e^{i(q-k+\hat{q})/k_*}}{q-k+\hat{q}}\right]\, 
\ea
where $\hat{q}=\sqrt{q^2+k^2+2kq\mu}$, $\mu_{\rm min}={\rm Max}\{-1,-\frac{|k_{\rm min}^2-q^2-k^2|}{2kq}\}$, and we take the normalization $ {\mathcal{B}}_{\rm GIS} \, =6/5$. The main characteristics of  $\mathcal{F}^{(3)}_{GIS}$ are listed below and illustrated in Figures \ref{fig:CurlyFcontributions} and \ref{fig:FofM}.

\begin{itemize}
\item{\bf $\mathcal{F}^{(3)}_{\rm GIS}$ depends strongly on nearly collinear configurations}. The largest contributions to the integrand come from triangles that are nearly collinear (elongated) and exactly collinear. This is shown in Figure \ref{fig:CurlyFcontributions}. The left panel shows the entire integrand, which peaks for $\mu$ closer to $\pm1$ than to 0. The right panel shows the contribution from the isosceles squeezed limit only (in this limit the shape of the integrand is very similar to the integrand for the exact local ansatz).
\item{\bf $\mathcal{F}^{(3)}_{\rm GIS}$ depends on the scale $k$}. The GIS shape is more divergent at small $k$ than the local ansatz. With the $f_i$ constant, $\mathcal{F}^{(3)}_{\rm GIS}$ can be nearly as divergent as $\frac{1}{k}$ while the local ansatz is constant. Constant contributions depending on $1/k_*$ level off the scale-dependence slightly. This is shown in the left panel of Figure \ref{fig:FofM}.
\item{\bf $\mathcal{F}^{(3)}_{GIS}$ depends on smoothing radius (mass)}. Since $\mathcal{F}^{(3)}_{GIS}$ is scale-dependent (the result is not invariant under rescalings $k\rightarrow\lambda k$), $\mathcal{F}^{(3)}_{GIS}$ depends on the smoothing scale $R$. This means that the amplitude of the non-Gaussian term $\Delta b_{NG}$ will depend differently on the mass of the object (galaxy or dark matter halo) whose power spectrum is considered than the Gaussian bias does. In addition, the last term in the square brackets in Eq.(\ref{eq:Deltabgeneric}) will contribute to the bias for the GIS shape. This is shown in the right panel of Figure \ref{fig:FofM}.
\item{\bf $\mathcal{F}^{(3)}_{GIS}$ has contributions from the real and imaginary parts of $f_i$}. Both panels of Figure \ref{fig:FofM} show $\mathcal{F}^{(3)}_{GIS}$ evaluated for $f_i=1$ (lines labeled ${\rm Re}(f_i)$) and $f_i=i$ (lines labeled ${\rm Im}(f_i)$). The contributions to the integrand from ${\rm Re}(f_i)$ come from both  isosceles-squeezed and near collinear-squeezed configurations, while those proportional to ${\rm Im}(f_i)$ are entirely from collinear-squeezed momenta.
\item{\bf $\mathcal{F}^{(3)}_{GIS}$ depends on $k_*$}. The contributions from squeezed collinear configurations depend on the scale $k_*$ and dominate $\mathcal{F}^{(3)}_{GIS}$. Notice however, that this does not imply the (physically unreasonable) result that non-Gaussianity increases arbitrarily for a long duration of inflation: here we have taken a simplified scenario where the coefficients $f_i$ are constants, which cannot hold over an arbitrarily long range of momenta. Figure \ref{fig:FofM} compares $k_*=10^{-5} Mpc^{-1}$ and $10^{-6} Mpc^{-1}$.
\end{itemize}

\begin{figure}[h]
\begin{center}
$\begin{array}{cc}
\includegraphics[width=0.5\textwidth,angle=0]{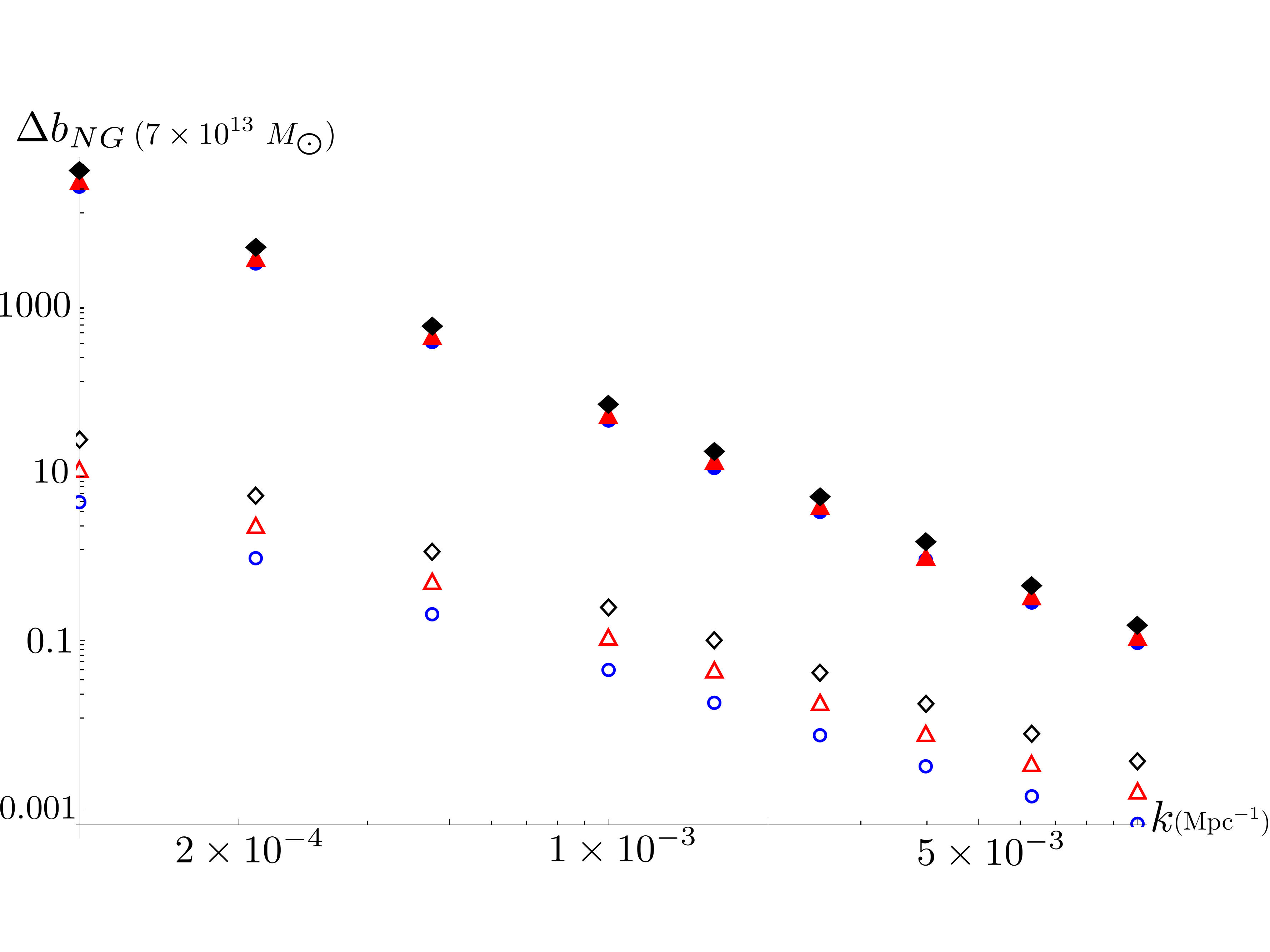} 
&
\includegraphics[width=0.5\textwidth,angle=0]{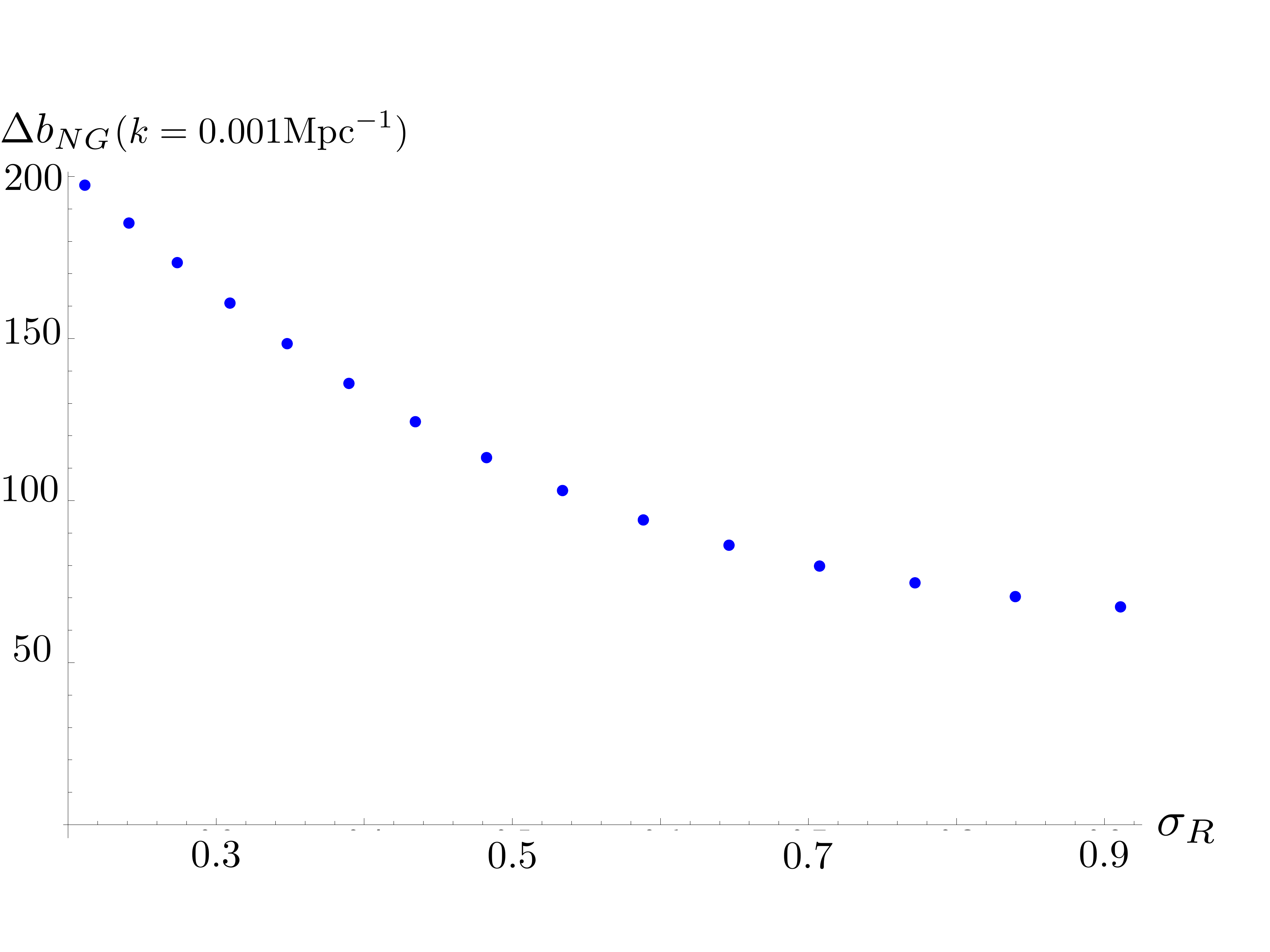} 
\end{array}$
\caption{The non-Gaussian bias from the Generalized Initial State with $\mathcal{B}_{\rm GIS}=\frac{6}{5}$ and $f_i$ real ($f_i$ imaginary contributes comparably). The left hand panel shows the result from GIS bispectrum (upper, filled points) compared to the local ansatz with $f_{NL}=1$ (lower, open points). Each case is computed for three masses: $7\times 10^{13} M_{\odot}$ (blue circles), $3\times 10^{14} M_{\odot}$ (red triangles), $1\times 10^{15} M_{\odot}$ (black diamond). The right hand panel shows the result from GIS bispectrum evaluated at $k=0.001 Mpc^{-1}$ as a function of $\sigma_R$. Smaller $\sigma_R$ is related to large scales and larger mass objects. See the bullet points in the text for more detailed explanations of these results.
\label{fig:MISbias}}
\end{center}
\end{figure}

The results for the non-Gaussian bias are shown in Figure \ref{fig:MISbias}. The qualitative features can be understood as consequences of the behavior of $\mathcal{F}^{(3)}_{GIS}$ discussed above. The most significant things to notice are that the bias for GIS (and constant $f_i$) can be as divergent as $1/k^3$ and can have a large amplitude on large scales, comparable to that from local $f_{NL}\sim6$, even if $\mathcal{B}_{\rm GIS}\sim (10^{-2})$.

\subsubsection{Contribution from the trispectrum} 

We can also study the contribution to the non-Gaussian bias coming from the primordial trispectrum (the four-point function in momentum space). The complete expression for the trispectrum for a generalized initial state is, however, more complicated than the expression for the bispectrum, and the numerical integrals involved in the computation of ${\mathcal{F}}_{\rm GIS}^{(4)}$ are more challenging. On the other hand, if we take into account that most of the contribution to the bias comes from the squeezed limit of primordial non-Gaussianity, we can obtain a good estimate by analyzing the limit of the trispectrum where one of the momenta is much smaller that the other three. This limit has been analyzed in \cite{Agullo:2011aa} and the conclusions are in parallel to those obtained for the bispectrum. Namely, the trispectrum shows a significant enhancement in the squeezed configurations that can be larger than the vacuum prediction by a factor $10^6$. This happens for squeezed configurations in which the three large momenta form a flattened triangle, the so called squeezed-flattened configurations for which $k_1\approx k_2\approx k_4/2 \gg k_3$ (although other squeezed configurations with similar enhancements may exist). It is then interesting to investigate the impact on the non-Gaussian bias.  In the squeezed-flattened limit, the four-point function takes the form \cite{Agullo:2011aa}

\be \langle \zeta_{\vec{k}_1}\zeta_{\vec{k}_2}\zeta_{\vec{k}_3}\zeta_{\vec{k}_4}\rangle=(2 \pi)^3\,  \delta_D(\sum_a \vec{k}_a) \,  T_{\zeta}(\vec{k}_1,\vec{k}_2,\vec{k}_3,\vec{k}_4)\ , \ee 
with 

 \be \label{result} T_{\zeta}(\vec{k}_1,\vec{k}_2,\vec{k}_3,\vec{k}_4)= \ \hat g_{NL} \ P_{\zeta}(k_1)  P_{\zeta}(k_3) P_{\zeta}(k_4) \, , \ee
where 

\be \label{tau1} \hat g_{NL}= 32 \, \epsilon \left(\frac{k_1}{k_3}\right)^{2} f_{\alpha,\beta} \cos{\theta} .\ee
In this expression, $\theta$ is the angle between the vectors $\vec{k}_3$ and $\vec{k}_4$, and $ f_{\alpha,\beta} $ contains the information about the initial state.  When the average number of initial quanta in the observable modes is of order one or greater, $f_{\alpha,\beta}$ is generally of order one.

The important point here is that when the above expression is employed to compute ${\mathcal{F}}_{\rm GIS}^{(4)}$, the integration over all squeezed-flattened configurations vanishes, due to the presence of  $\cos{\theta}$. Therefore, even when those configurations are enhanced compared to the vacuum case, they do not produce a significant contribution to the non-Gaussian bias. We can not discard, however, that other configurations produce a significant contribution, and a more detailed analysis is needed. 

\subsection{The skewness}
\begin{figure}[h]
\begin{center}
$\begin{array}{cc}
\includegraphics[width=0.5\textwidth,angle=0]{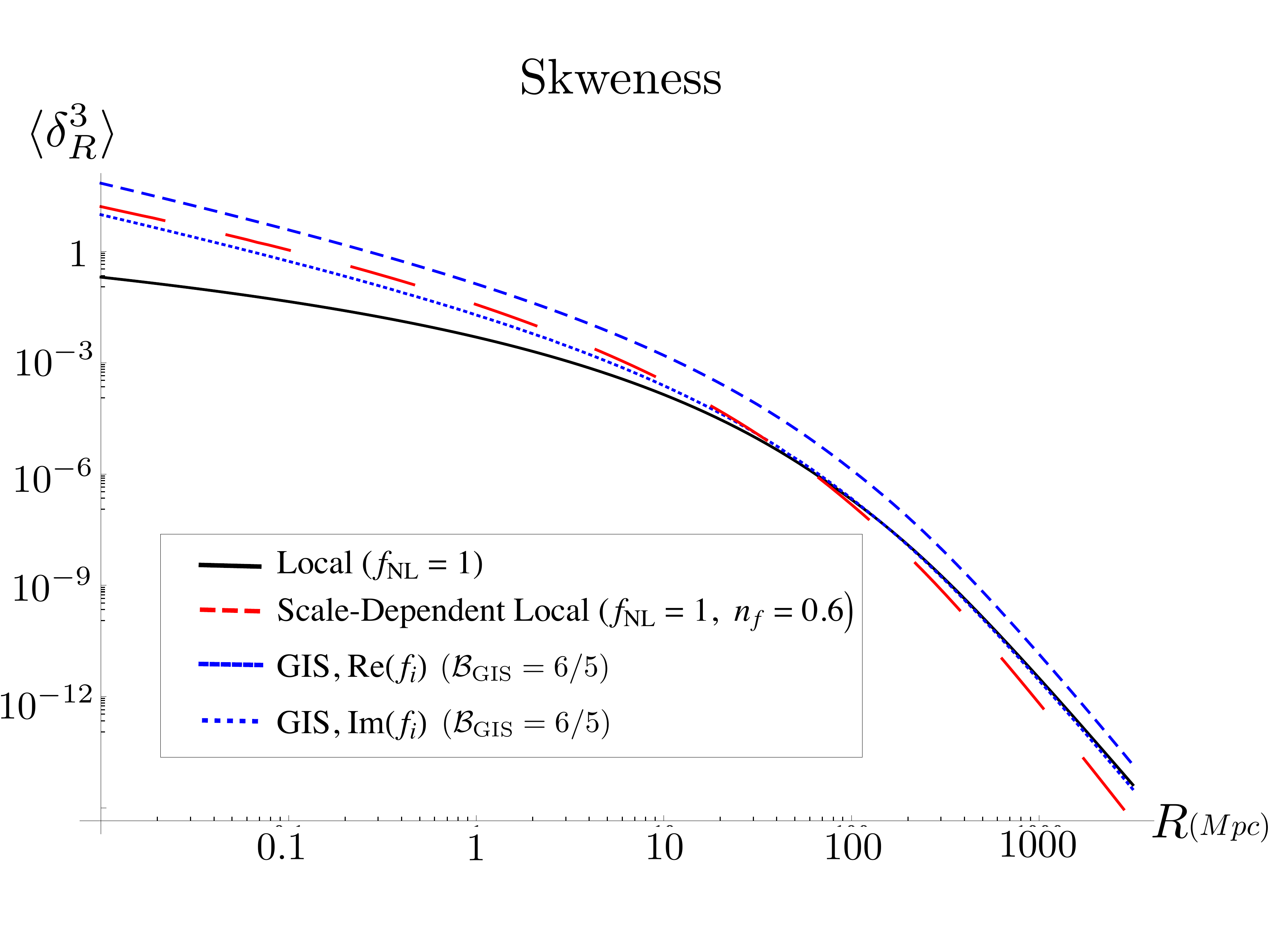} 
&
\includegraphics[width=0.5\textwidth,angle=0]{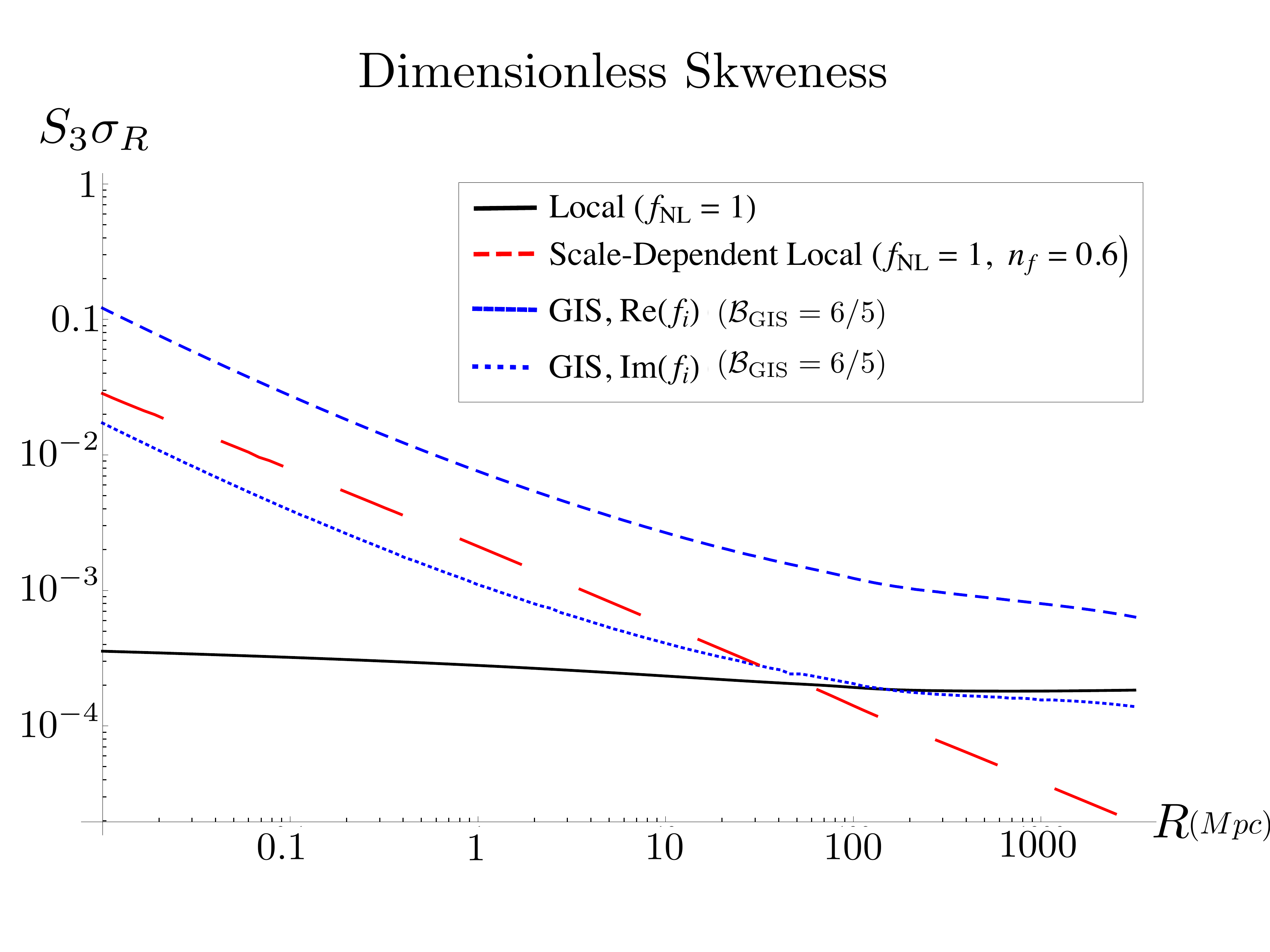} 
\end{array}$
\caption{Comparing the size of the skewness for the usual local ansatz (solid black) and the Generalized Initial State (blue short dashed). We take $f_i=1+i$ and separately plot the contributions proportional to the real and imaginary parts of the $f_i$. For reference we also show a scale-dependent local ansatz, where $f_{NL}=f_{NL}(k_p)(k/k_p)^{n_f}$ (the red, long-dashed line). The right panel shows the dimensionless skewness for the same bispectra.}
\label{fig:CompareDelta3}
\end{center}
\end{figure}

The skewness is useful for getting a sense of the overall level of non-Gaussianity since it integrates over the full bispectrum. It is also what appears in the non-Gaussian mass function. The smoothed 3-point function is
\ba
\label{eq:smooth3point}
\langle\delta^3_R\rangle&=&\int\frac{d^3k_1}{(2\pi)^3}\int\frac{d^3k_2}{(2\pi)^3}\int\frac{d^3k_3}{(2\pi)^3}\,M_R(k_1)M_R(k_2)M_R(k_3)\langle \zeta_{\vec{k}_1}\zeta_{\vec{k}_2}\zeta_{\vec{k}_3}\rangle \, .
\ea
For a Generalized Initial State with the $f_i$ constant that is
\ba
\langle\delta^3_R\rangle&=&\frac{3}{2}\left(\frac{3}{5}\right)^3\mathcal{B}_{\rm GIS}\int_{k_{\rm min}}^{\infty} dk_1\int_{k_{\rm min}}^{\infty} dk_2\int_{\mu_{min}(k_1,k_2)}^{1}d\mu\; \\\nonumber
&&\times M_R(k_1)M_R(k_2)M_R(\hat{k})\mathcal{P}_{\zeta}(k_1)\mathcal{P}_{\zeta}(\hat{k})\frac{k_1}{k_2\hat{k}}\\\nonumber
&&\times\; \textrm{Re}\left\{f_t\frac{1-e^{i(k_1+k_2+\hat{k})/k_*}}{k_1+k_2+\hat{k}}+f_1\frac{1-e^{i(-k_1+k_2+\hat{k})/k_*}}{-k_1+k_2+\hat{k}}\right.\\\nonumber
&&\left.+f_2\frac{1-e^{i(k_1-k_2+\hat{k})/k_*}}{k_1-k_2+\hat{k}}+f_3\frac{1-e^{i(k_1+k_2-\hat{k})/k_*}}{k_1+k_2-\hat{k}}\right\} \, ;\\\nonumber
\hat{k}&=&\sqrt{k_1^2+k_2^2+2k_1k_2\mu} \, .
\ea
We again take the normalization $\mathcal{B}_{\rm GIS}=\frac{6}{5}$ and the simplified parameter case $f_i=1+i$. We use $k_*=10^{-5} {\rm Mpc}^{-1}$ so that all currently observed CMB modes satisfy $k_{\rm obs}\geq10k_*\equiv k_{\rm min}$. It's also interesting to compare the shape as a function of smoothing scale (mass) a little more carefully by plotting the dimensionless skewness, which is nearly scale-independent for the local ansatz:
\be
S_{3,R}\sigma_R=\frac{\langle\delta^3_R\rangle}{\langle\delta^2_R\rangle^{3/2}} \, .
\ee
The results for the GIS shape and the local bispectrum with $f_{NL}=1$ are plotted in Figure \ref{fig:CompareDelta3}. 

These results for the skewness suggest that we could profitably use the already calculated Minkowski functional constraints from the CMB as a first pass at constraining the GIS non-Gaussianity. Minkowski functionals, reported for the local ansatz, restrict $-70<f_{NL}<91$ at 95\% confidence level \cite{Hikage:2008gy,Natoli:2009wk}. (For comparison, constraints from WMAP \cite{Komatsu:2010fb} and the local ansatz bispectrum are $-10<f_{NL}<74$.) Finally, since the GIS non-Gaussianity is scale-dependent, a combination of CMB and LSS constraints, including cluster counts, would ultimately constrain this physics in complementary ways \cite{LoVerde:2007ri, Sefusatti:2009xu, Becker:2010hx, Shandera:2010ei}. 

\section{Conclusions}
\label{sec:conclusions}
In this paper we have analyzed the consequences of non-Gaussianity from single field inflation with a generalized initial state. We have especially emphasized the strongly scale-dependent term in the halo bias. The strongest scale dependence in the bias corresponds to a choice for the initial state that is nearly scale invariant (over a finite range of momenta). Near scale-invariance of the initial state is necessary to satisfy constraints from the spectral index when deviations from Bunch-Davies are non-negligible.

Perhaps the most interesting implication of our result is that it strengthens the case for a more general analysis of existing Large Scale Structure data and bias, which has already been used to significantly constrain non-Gaussianity of exactly the local type. That constraint is remarkable, but the data are powerful enough to do even more. Only a very small fraction of inflation scenarios predict non-Gaussianity of exactly the local type, so it is useful to characterize the non-Gaussian bias in the most general way the data allows. Our work here, together with the work on quasi-single field inflation \cite{Chen:2009zp,Chen:2009we} and genuinely two field models \cite{Shandera:2010ei} suggest that it would be very profitable to constrain non-Gaussian bias in terms of the two parameter family 
\be
\Delta b_{NG}\propto\frac{f_{NL}^{\rm eff}(M)}{k^\alpha}
\ee
allowing $\alpha$ to be a continuous parameter in at least the range $\alpha\leq3$.

It is perhaps useful to note that, at least for the purposes of large scale structure observables, the qualitatively important features of single field Generalized Initial State, quasi-single field, and multi-field models can all be captured by the generalized local ansatz introduced in \cite{Shandera:2010ei}:
\be
B_{\Phi}(\vec{k}_1,\vec{k}_2,\vec{k}_3)=\xi_{s}(k_2)\xi_{m}(k_1)\xi_{m}(k_3)P_{\Phi}(k_1)P_{\Phi}(k_3)+5\:{\rm
  perm}\;.  
\ee
where the $\xi_{s,m}$ are allowed to be independent power law functions of the momenta (compared to some pivot point $k_p$):
\begin{equation}
\xi_{s,m}(k)=\xi_{s,m}(k_p)\left(\frac{k}{k_p}\right)^{n^{(s),(m)}_f}
\label{eq:xi_powerlaw}
\end{equation}
For example, we see from Eq.(\ref{eq:BwithLargeN}), which was written for $k_1=k_2$, that a Generalized Initial State populated by 
\be N_k=N_0 \left(\frac{k}{k_*}\right)^{-\delta} \, , \ee
has $n^{(s)}_f=2-2\delta$ and $n^{(m)}_f=\delta-1$. Physically reasonable versions of any of the models listed above will of course allow only restricted versions of the generalized local ansatz (eg, multi-field models typically have the $n^{(s),(m)}_f$ of order slow-roll parameters). Still, the ansatz may be useful for phenomenological modeling of generic effects of correlations between long and short wavelength modes.

\acknowledgments
It is a pleasure to thank Louis Leblond and Abhay Ashtekar for helpful discussions and comments on the manuscript.  We also thank J. Ganc and E. Komatsu for friendly coordination over the course of this work. This work has been supported by NSF grant PHY-0854743 and by the Eberly Research Funds of The Pennsylvania State University.
The Institute for Gravitation and the Cosmos is supported by the Eberly College of Science and the Office of the Senior Vice President for Research at the Pennsylvania State University.

\appendix

\section{Bispectrum from a Generalized Initial State}

Here we write the complete expression of the inflationary bispectrum obtained from allowing a more generic quantum state for the scalar perturbations. We consider states specified by a Bogoliubov transformation of the Bunch-Davies vacuum. 
The computation of the bispectrum $B_{\rm GIS}$ follows the same steps as the computation using the vacuum states \cite{Maldacena:2002vr},  with the vacuum mode functions substituted by the Bogoliubov rotated mode functions $\bar{\zeta}_{k}(\tau)$ (see Eq.(\ref{modes})). We do not reproduce here the details of the derivation, and we refer the reader to the literature \cite{Holman:2007na} \cite{Agullo:2010ws} \cite{Ganc:2011dy}. The result is given by
\ba \label{fullbisp}
& &B_{\rm GIS}(k_1,k_2,k_3) = P_{\zeta}(k_1)P_{\zeta}(k_2)  \Big\{ \frac{1}{2} \left(3 \epsilon-2\eta +\epsilon \,  \frac{k_1^2+k_2^2}{k_3^2} \right) + \\\nonumber
&+&\, 4 \epsilon \, \frac{k_1^2k_2^2}{k_3^3} \ \textrm{Re}\left[f_t\frac{1-e^{ik_t/k_*}}{k_t}+f_{1}\frac{1-e^{i\tilde{k}_1/k_*}}{\tilde{k}_1}+f_{2}\frac{1-e^{i\tilde{k}_2/k_*}}{\tilde{k}_2}+f_{3}\frac{1-e^{i\tilde{k}_3/k_*}}{\tilde{k}_3} \right]  \Big\}  +\textrm{2 perm.}\, ,\ea
where $k_t=k_1+k_2+k_3$, $\tilde{k}_i=k_t-2 k_i$,  
and $k_*^{-1}\equiv\tau_0$ characterizes the value of the conformal time at which the initial conditions for inflation are specified. The physical condition that the observable modes in our present universe were deeply inside the Hubble radius at the onset of inflation translates into the condition $k_i\gg k_*$, for $i=1,2,3$. In the previous equation we also have
$$ P_{\zeta}(k)=|\bar{\zeta}_k|^2=\frac{1}{2 \epsilon M_P^2} \frac{H^2}{2 k^3} |\alpha_{k}+\beta_{k}|^2\ . $$
 $$f_t=\frac{1}{\prod_{i=1}^2 |\alpha_{k_i}+\beta_{k_i}|^2}\Big[\prod_{i=1}^3(\alpha_{k_i}+\beta_{k_i})(\alpha^*_{k_1} \alpha^*_{k_2} \alpha^*_{k_3})- \prod_{i=1}^3(\alpha^*_{k_i}+\beta^*_{k_i})(\beta_{k_1} \beta_{k_2} \beta_{k_3})\Big]\ ,$$ $$f_{1}=\frac{1}{\prod_{i=1}^2 |\alpha_{k_i}+\beta_{k_i}|^2}\Big[\prod_{i=1}^3(\alpha_{k_i}+\beta_{k_i})(\beta^*_{k_1} \alpha^*_{k_2} \alpha^*_{k_3})-\prod_{i=1}^3 (\alpha^*_{k_i}+\beta^*_{k_i})(\alpha_{k_1} \beta_{k_2} \beta_{k_3})\Big] \ , $$  $$f_{ 2}=\frac{1}{\prod_{i=1}^2 |\alpha_{k_i}+\beta_{k_i}|^2}\Big[\prod_{i=1}^3(\alpha_{k_i}+\beta_{k_i})(\alpha^*_{k_1} \beta^*_{k_2} \alpha^*_{k_3})- \prod_{i=1}^3 (\alpha^*_{k_i}+\beta^*_{k_i})(\beta{k_1} \alpha_{k_2} \beta_{k_3})\Big] \ , $$   $$f_{3}=\frac{1}{\prod_{i=1}^2 |\alpha_{k_i}+\beta_{k_i}|^2}\Big[\prod_{i=1}^3 (\alpha_{k_i}+\beta_{k_i})(\alpha^*_{k_1} \alpha^*_{k_2} \beta^*_{k_3})-\prod_{i=1}^3 (\alpha^*_{k_i}+\beta^*_{k_i})(\beta_{k_1} \beta_{k_2} \alpha_{k_3})\Big] \ , $$ 
If there are no particles present in the initial state, then  $\beta_{k}= 0$, in which case $f_t=1$, $f_1=f_2=f_3=0$, and expression (\ref{fullbisp}) reproduces the well known slow-roll result obtained in \cite{Maldacena:2002vr}. 

\bibliography{GISbiasRevised}{}
\bibliographystyle{JHEP}

\end{document}